\documentclass{article}
\usepackage{adjustbox}
\usepackage{arxiv}
\usepackage[utf8]{inputenc} 
\usepackage[T1]{fontenc}    
\usepackage[colorlinks=true]{hyperref}       
\usepackage{url}            
\usepackage{booktabs}       
\usepackage{amsfonts}       
\usepackage{nicefrac}       
\usepackage{microtype}      
\usepackage{lipsum}
\usepackage{fancyhdr}       
\usepackage{graphicx}       
\graphicspath{{media/}}     
\usepackage{amssymb}
\usepackage{amsmath}
\usepackage{mathtools}
\usepackage{placeins}
\usepackage[numbers]{natbib}
\usepackage{caption}
\usepackage{verbatim}
\usepackage{rotating}
\usepackage[utf8]{inputenc}
\usepackage{pgfplots}
\usepackage{appendix}
\usepgfplotslibrary{groupplots}
\pdfoutput=1

\newcommand{\vx}{\boldsymbol{x}}
\newcommand{\vz}{\boldsymbol{z}}
\newcommand{\vu}{\boldsymbol{u}}

\newcommand{\vy}{\boldsymbol{y}}

\newcommand{\ve}[1]{\boldsymbol{#1}}
\renewcommand{\shorttitle}{Bayesian tomography using polynomial chaos expansion and deep generative networks}

\author{
  Giovanni Angelo Meles\\
    Institute of Earth Sciences \\
      University of Lausanne \\
      Switzerland\\
    \texttt{Giovanni.Meles@unil.ch} \\
   \And
  Macarena Amaya \\
    Institute of Earth Sciences \\
      University of Lausanne \\
      Switzerland\\
  \texttt{Macarena.Amaya@unil.ch} \\
  \And
  Shiran Levy \\
    Institute of Earth Sciences \\
      University of Lausanne \\
      Switzerland\\
  \texttt{Shiran.Levy@unil.ch} \\
  \And
  Stefano Marelli \\
  Institute of Structural Engineering \\
  ETH Zurich \\
  Switzerland\\
  \texttt{Marelli@ibk.baug.ethz.ch} \\
       \And
     Niklas Linde \\
    Institute of Earth Sciences \\
      University of Lausanne \\
      Switzerland\\
    \texttt{Niklas.Linde@unil.ch} \\
}

\begin{document}

\title{Surrogate-based Bayesian traveltime tomography with geologically-complex priors}
\date{\today}
\maketitle
\begin{abstract}

Implementations of Markov chain Monte Carlo (MCMC) methods need to confront two fundamental challenges: accurate representation of prior information and efficient evaluation of likelihoods.
Principal component analysis (PCA) and related techniques can in some cases facilitate the definition and sampling of the prior distribution, as well as the training of accurate surrogate models, using for instance, polynomial chaos expansion (PCE). However, complex geological priors with sharp contrasts necessitate more complex dimensionality-reduction techniques, such as, deep generative models (DGMs).
By sampling a low-dimensional prior probability distribution defined in the low-dimensional latent space of such a model, it becomes possible to efficiently sample the physical domain at the price of a generator that is typically highly non-linear. Training a surrogate that is capable of capturing intricate non-linear relationships between latent parameters and outputs of forward modeling presents a notable challenge. Indeed, while PCE models provide high accuracy when the input-output relationship can be effectively approximated by relatively low-degree multivariate polynomials, this condition is typically not met when employing latent variables derived from DGMs.
In this contribution, we present a strategy combining the excellent reconstruction performances of a variational autoencoder (VAE) with the accuracy of PCA-PCE surrogate modeling in the context of Bayesian ground penetrating radar (GPR) traveltime tomography. 
Within the MCMC process, the parametrization of the VAE is leveraged for prior exploration and sample proposals. Concurrently, surrogate modeling is conducted using PCE, which operates on either globally or locally defined principal components  of the VAE samples under examination. Our methodology is exemplified using channelized subsurface structures, providing accurate reconstructions and significant speed-ups, particularly when the computation of the full-physics forward model is costly.

\end{abstract}

\section{Introduction}

Bayesian inversion methods can account for data and modeling uncertainties as well as prior knowledge, thus,
representing an attractive approach for tomography and its uncertainty quantification.
Nevertheless, the difficulties in specifying appropriate prior distributions and the high computational burden associated with repeated forward model evaluations  often hinder proper implementations of Bayesian tomography 
\citep{chipman2001practical}. 
In geophysical settings, prior distributions have traditionally been specified by assuming the subsurface to be represented by a Gaussian random field. More advanced options are made possible by relying on the information content of training images (TI), that is, large gridded 2-D or 3-D unconditional representations of the expected target spatial field that can be either continuous or categorical \citep{mariethoz2014multiple,laloy2017inversion, laloy2018training}. Bayesian inversion needs not only a parameterization of the prior that accurately represents the prior information, but also one that is easy to manipulate and one for which small perturbations lead to comparatively small changes in the data response. In many cases, it is advantageous to rely on another parameterization than the one used in physics-based modeling and visualization, that is, typically a pixel-based parameterization. Physical media are then associated with points in $\mathbb{R}^N$, where $N$ is the number of elements in the corresponding pixel-based representation. While allowing easy implementation of forward modeling schemes (e.g., Finite Difference (FD) based on partial differential equations), pixel-based $N$-dimensional parametrizations are often not suitable to effectively parametrize the prior distribution, as $N$ can be very large.
When prior knowledge suggests constrained spatial patterns, such as covariance or connected spatial structures, the prior-compatible models populate manifolds embedded in $\mathbb{R}^N$.
If this manifold can be locally assimilated to a subset of 
$\mathbb{R}^M$, with $M \ll N$, the prior distribution reduces to a function of $M$ variables only, which leads to lower-dimensional inverse problems.

Various approaches can be employed to achieve manifold identification through dimensionality reduction. Among these techniques, principal component analysis (PCA) and related methods are the most common \textit{linear} dimensionality reduction methods \citep{boutsidis2008unsupervised,jolliffe2016principal}.
Based on a data set of prior model realizations and the eigenvalues/eigenvectors of the corresponding covariance matrix, PCA provides optimal $M$-dimensional representations in terms of uncorrelated variables that retain as much of the sample variance as possible. 
PCA has found widespread application in geophysics, in both deterministic and stochastic inversion algorithms, with recent advancements offering the potential to reconstruct even discontinuous structures \citep{reynolds1996reparameterization,giannakis2021fractal,meles2022bayesian,thibaut2021new}.
In the context of complex geological media with discrete interfaces \citep{strebelle2002conditional,zahner2016image,laloy2018training} investigated in this study, PCA-based methods are insufficient for inversion. 
Promising alternatives are offered by deep generative models (DGM) that learn the underlying input distribution and generate synthetic samples that closely resemble a provided dataset. For instance, Variational Autoencoders (VAEs) encode data patterns into a compact latent space for both data reconstruction and generation. Generative Adversarial Networks (GANs) employ adversarial training to create synthetic output that closely resembles real reference data. In contrast, Diffusion Models are parameterized Markov chains, trained using variational inference, to generate samples that converge to match the underlying data distribution 
\citep{kingma2013auto,jetchev2016texture,goodfellow2020generative,ho2020denoising}. 
In the context of Bayesian inversion, DGMs possess a crucial property: they generate realizations that exhibit patterns consistent with the TI when sampling an uncorrelated standard normal or uniform distributed latent space \citep{laloy2017inversion}. The incorporation of such a low-dimensional parameterization representing the prior distribution simplifies the sampling process, thus making DGMs well-suited for MCMC schemes.

Employing an effective parametrization for the prior (e.g., via PCA or DGMs) does not alleviate the computational cost associated with repeated forward modeling, which can often limit practical application of MCMC schemes.
To address this concern, a potential solution is to explore surrogate modeling. 
Various classes of surrogate models, such as those based on Gaussian process models or Kriging
\citep{sacks1989designs,rasmussen2003gaussian} and polynomial chaos expansions (PCEs) \citep{xiu2002wiener,blatman2011adaptive}, can be employed in low-dimensional Bayesian inverse problems \citep{nagel2019bayesian,higdon2015bayesian,marzouk2009stochastic,marzouk2007stochastic,wagner2020bayesian,wagner2021bayesian}.

Recently, \citet{meles2022bayesian}
presented a Bayesian framework that employs a shared PCA-based parameterization for characterizing prior information and modeling travel times through PCE.
In this paper, we show that the direct application of the methodology introduced in \citet{meles2022bayesian}
produces sub-optimal results when input is parametrized in terms of latent variables associated with a DGM.
Indeed, a prerequisite for the successful  application of the strategy introduced by \citet{meles2022bayesian} is that the parametrization chosen to describe and explore the prior allows for implementation of accurate surrogate modeling of the physical data used in the inversion. However, the relationship between the parametrization provided by DGMs and the corresponding physical output is typically highly non-linear, and training a surrogate capturing such complex relationship can be extremely challenging or infeasible. 

We here present a strategy that allows the use of a DGM parametrization to define the prior distribution and explore the posterior distribution while still enabling accurate surrogate modeling. We achieve this goal by employing for forward modeling with either global or local PCA decompositions of the input generated by the underlying DGM during the MCMC process. Through the decoupling of the MCMC parameterizations and forward modeling parametrizations, we preserve the advantageous prior representation capabilities of DGMs while simultaneously achieving precise PCA-based surrogate modeling, thereby significantly speeding up forward calculations.

\section{Methodology}
\subsection{Bayesian inversion}
\label{sBayes}
Forward models are mathematical tools that quantitatively evaluate the outcome of physical experiments. We refer to the relationship between input parameters and output values as the 'forward problem':
\begin{equation}
 \mathcal{F}(\vu) = \vy + \epsilon.
 \label{forward}
\end{equation}
Here, $\mathcal{F}$, $\vu$, $\vy$ and $\epsilon$ stand for the physical law or forward operator,
the input parameters, typically representing local media properties, the output and a noise term, respectively.
The goal of the 'inverse problem' is to infer properties of $\vu$ conditioned by the data $\vy$ while taking into account any available prior information about $\vu$.
A general solution to this problem can be expressed in terms of the posterior distribution defined over the input domain by Bayes' theorem:
\begin{equation}
 P(\vu|\vy)= \frac{P(\vy|\vu)P(\vu)}{P(\vy)}
 \label{Bayes}=\frac{L(\vu)P(\vu)}{P(\vy)}.
\end{equation}
Here, $P(\vu|\vy)$ is the posterior distribution of the input parameter $\vu$ given the data $\vy$, $P(\vy|\vu)$ (also indicated as $L(\vu)$ and known as 'the likelihood') is the probability of observing the data $\vy$ given the input parameter $\vu$, while $P(\vu)$  and $P(\vy)$ are the prior distribution in the input parameter domain and the marginalized likelihood with respect to the input parameters (also known as evidence), respectively. Note that throughout this paper, we adhere to the common formalism employed in Geophysics, utilizing the same symbol to denote both individual instances of a random variable and the random variable itself  \citep{tarantola2005inverse,aster2018parameter}. In practical applications, Eq. ~\eqref{Bayes} is seldom used when solving inverse problems as the computation of the evidence is in most cases very expensive. However,  since the numerator of the right-hand side of Eq. ~\eqref{Bayes}, that is, $L(\vu)P(\vu) $, is proportional to the posterior distribution, one can use a Markov chain Monte Carlo (MCMC) methods to draw samples proportionally from $P(\vu|\vy)$ without considering the evidence \citep{hastings1970monte}. 
However, computing $L(\vu)$ requires the solution of a forward problem, which can be demanding in Bayesian inversions as this evaluation needs to be repeated many times. In the following sections we discuss how this  problem can be approached by using a latent representation and surrogate modeling to evaluate $P(\vu)$ and approximate $L(\vu)$, respectively.

\label{metodo}
\subsection{Bayesian inversion in latent spaces}
Parametrizations are particularly well suited for surrogate modeling when they can easily encode the prior distribution and  effectively simplify the input-output relationship within the problem under investigation.  
\citet{meles2022bayesian} used variables defined in terms of Principal Components to (a) represent the prior distribution related to a random Gaussian field on a low-dimensional manifold \textit{and} (b) learn an accurate surrogate to compute the forward problem. However, it is not generally granted that a parameterization can achieve both (a) and (b).
For representing the prior distribution, we can utilize manifold identification 
using DGMs. This involves utilizing a DGM to characterize a latent space using a set of coordinates (here indicated as $\vz$) and a statistical distribution defined prior to the training. The DGM allows mapping between this  latent space to the physical space through a deep learning decoder, denoted here as $\mathcal{G}_{DGM}$. For a given random realization $\vz$, the decoder operation $\mathcal{G}_{DGM}(\vz)$ produces an output $\vu$ in the physical space that adheres to the characteristics of the prior distribution. 
The use of this new set of coordinates ${\vz}$ casts the inverse problem on the latent manifold as:
\begin{equation}
 P(\vz|\vy)= \frac{P(\vy|\vz)P(\vz)}{P(\vy)}.
 \label{BayesReduced}
\end{equation}
While formally identical to Eq. ~\eqref{Bayes}, Eq. ~\eqref{BayesReduced} involves significant advantages.
For this class of coordinates,  not only $\vz$ is typically low-dimensional (consisting typically of a few tens of variables instead of many thousands of variables) but we can also design the corresponding statistical prior distribution $P(\vz)$ as desired. In this case, we impose during training that $P(\vz)$ is a multivariate standard Gaussian distribution.

\subsection{Decoupling of inversion and modeling domains in  MCMC inversions}
\label{effective}
We now rewrite the forward problem in Eq. ~\eqref{forward} using the new coordinates:
\begin{equation}
 \mathcal{M}_{DGM}({\vz}) = {\vy} +\epsilon,
 \label{forwardcoordinates_01}
\end{equation}
where $\mathcal{M}_{DGM}=\mathcal{F\circ G}_{DGM}$,  $\circ$ stands for function composition, and we assume no error induced by the DGM dimensionality reduction.
The complexity and non-linearity of $\mathcal{G}_{DGM}$ imply that the forward operator $\mathcal{M}_{DGM}$ will exhibit considerable irregularity, making it difficult to learn a surrogate model. Consequently, we investigate alternative approaches that avoids using the latent parametrization for surrogate modeling while retaining it for the prior representation. Building upon \citet{meles2022bayesian}, we explore surrogate modeling based on PCA-inputs spanning the \textit{Global} spatial extent of the input. 
Without any loss of generality, we consider a complete set of Global Principal Components (in the following, GPCs) for realizations of the DGM (implemented via $\mathcal{G}_{GPC} (\vx^{full}_{GPC})= \mathcal{G}_{DGM}({\vz}) = \vu$) and rewrite Eq. ~\eqref{forward} as:
\begin{equation}
 \mathcal{M}_{GPC}(\vx^{full}_{GPC}) = {\vy} +\epsilon,
 \label{forwardcoordinates_02}
\end{equation}
where $\mathcal{G}_{GPC}$, and $\mathcal{M}_{GPC}$  are the physical distribution and the model associated with the GPCs and therefore  $\mathcal{M}_{GPC}=\mathcal{F\circ G}_{GPC}$.
We will show in what follows that the linear relationship $\mathcal{G}_{GPC} (\vx^{full}_{GPC})=\vu$ can make it suitable for implementing a surrogate of $\mathcal{M}_{GPC}$, provided that the input and the model can be faithfully represented as operating on an effective $M$-dimensional truncated subset $\vx_{GPC}$ of the new coordinates $\vx^{full}_{GPC}$, that is:
\begin{equation}
\mathcal\mathcal{G}_{GPC} (\vx_{GPC}) \approx \vu  \implies \mathcal{M}_{GPC}({\vx}_{GPC}) = {\vy} +\hat{\epsilon},
 \label{forwardreduced}
\end{equation}
where $\hat{\epsilon}$ is a term including both observational noise and modeling errors related to the projection on the subset represented by $\vx_{GPC}$. 

The argument presented above relies on the \textit{weak} hypothesis that \textit{Global} proximity in the  input domain leads to proximity in the output domain $\vu$. Critically, when the output functionally depends mainly on a subset $L$ of the entire domain of $\vu$, proximity in the output domain can be attained by approximating the input within this \textit{Local} region. Based on this \textit{strong} physically-informed assumption and following the argument discussed above, we can achieve this goal by means of a Local PCA decomposition restricted to $L$:
\begin{equation}
\mathcal\mathcal{G}_{LPC} (\vx_{LPC})\upharpoonright_L \approx \vu\upharpoonright_L \implies  \mathcal{M}_{LPC}({\vx}_{LPC}) = {\vy} +\hat{\epsilon},
 \label{Lforwardreduced}
\end{equation}
where ${LPC}$ refers to the use of Local Principal Components (in the following LPCs) and $\upharpoonright_L$ restricts the validity of these relationships to the subset $L$. However, because the area spanned by LPCs is smaller than that of GPCs, we can expect to need fewer ${LPC}$s than ${GPC}$s  to achieve a satisfactory approximation in the output domain. Note that such a change of coordinates is not invertible even if a complete set of LPCs is employed.
\label{PCEtheory}

Generally speaking, a surrogate model 
is a function that seeks to emulate the behaviour of an expensive forward model at negligible computational cost per run. Clearly, the function a forward solver has to model depends on the set of coordinates used to represent the input. 
For simplicity, we discuss here a  surrogate for $\mathcal{M}_{GPC}({\vx}_{GPC})$, namely $\tilde{\mathcal{M}}_{GPC}$ satisfying:
\begin{equation}
\tilde{\mathcal{M}}_{GPC}(\vx_{GPC}) \approx \mathcal{M}_{GPC}(\vx_{GPC}).
 \label{surrogate}
\end{equation}
Once available, surrogate models can be used for likelihood evaluation in MCMC inversions, with a modified  covariance operator $\boldsymbol{C}_D = \boldsymbol{C}_d+\boldsymbol{C}_{Tapp}$ comprising the covariance matrices  $\boldsymbol{C}_d$ and $\boldsymbol{C}_{Tapp}$ accounting for  data uncertainty and  modeling error, respectively. In these cases the likelihood is expressed as:
      \begin{equation} 
      \label{likely}
   L(\vx_{GPC})=   \left( \frac{1}{2 \pi} \right)^{n / 2} |\boldsymbol{C_D}|^{-1/2} \mbox{exp} 
           \left[ -\frac{1}{2} (\mathcal{\tilde{M}}_{GPC}{(\vx_{GPC})}  - {\vy} )^T {\boldsymbol{C_D}}^{-1} (\mathcal{\tilde{M}}_{GPC}{(\vx_{GPC})}  - {\vy}) \right] \,.
\end{equation}
where $|\boldsymbol{C_D}|$ is the determinant of the covariance matrix $\boldsymbol{C_D}$ \citep{hansen2014accounting}.

\begin{figure}
\centering
\includegraphics[width=1\textwidth]{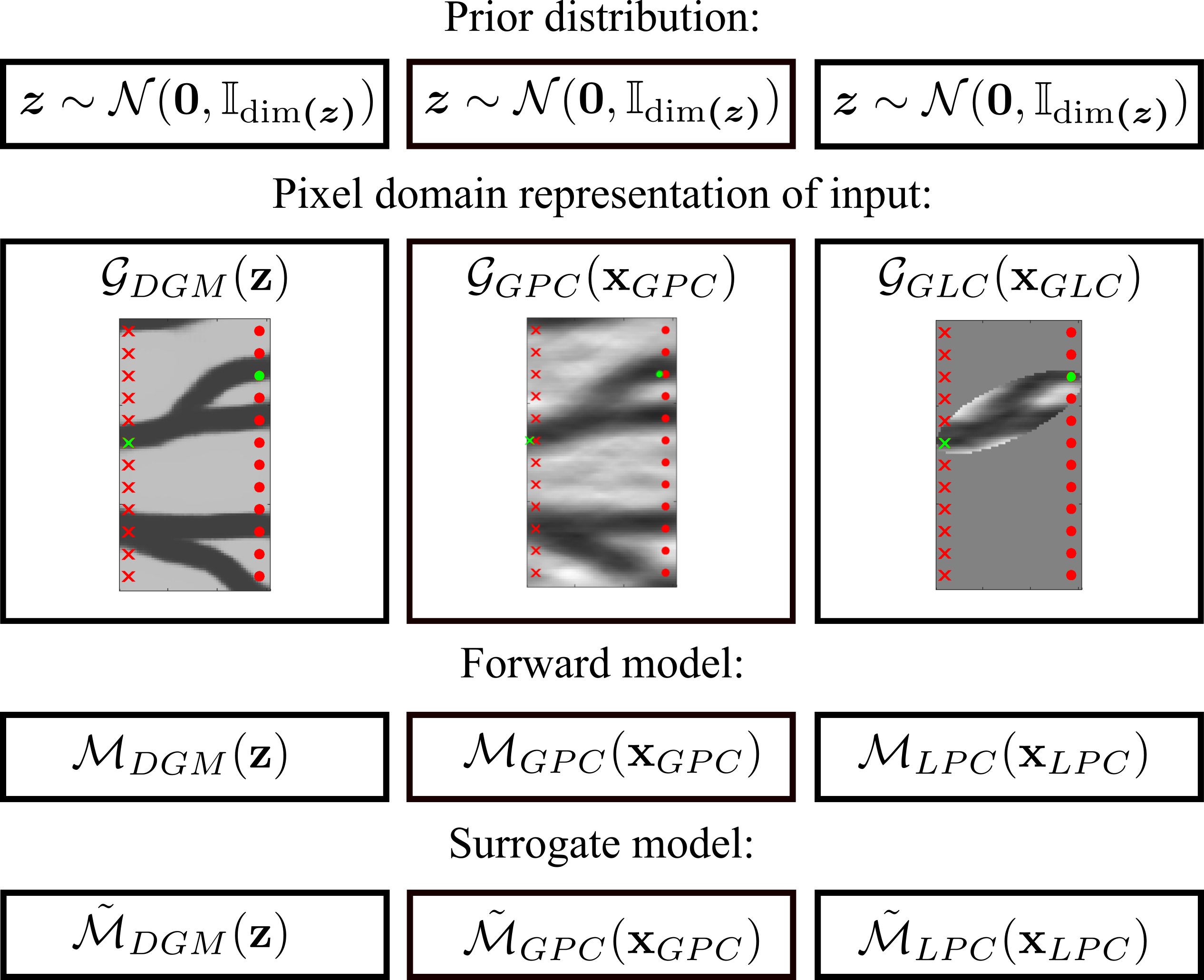}
\caption{\label{fig:DGM_DIAGRAM} Graphical schematic of the three strategies discussed in this manuscript. The prior distribution is  always parametrized by the latent space of the DGM, with $\vz \sim \mathcal{N}(0,\mathbb{I}_{dim(\vz)})$, while the input to the reference/surrogate model depends on the chosen strategy. 
When the DGM parametrization is chosen, draws of the latent variables are considered (left column). With the PCA strategies,  Globally or Locally defined G/LPCs are used for the Global and Local approach, respectively (central and right column). In this illustration, as we consider travel time modeling, the Local domain is defined in terms of fat-ray sensitivity kernels (for more details, please refer to section \ref{parametrization}
).
}
\end{figure}

There are various categories of surrogate models, such as Kriging and PCE, each with its own strengths and limitations. However, they all adhere to a fundamental principle: the more complex the input-output relationship, the higher the computational demands needed to build the surrogate model (e.g. in terms of training set). Furthermore, the efficiency of constructing a surrogate is significantly affected by the number of input parameters, and it can become unfeasible when the input dimensionality exceeds several tens of parameters, thus surrogate modeling often relies on some kind of dimensionality reduction. The dimensionality reduction step does not necessarily need to be invertible since what holds significance is the supervised performance, specifically the minimization of modeling error
\citep{lataniotis2020extending}. Surrogate models exhibit their peak potential when dealing with low-dimensional input spaces, provided that such simplicity does not entail a complex input-output relationship.

Based on this general considerations, we can qualitatively  anticipate the surrogate performance when operating on latent variables (i.e., $\tilde{\mathcal{M}}_{DGM}(\vz)$),  GPCs (i.e., $\tilde{\mathcal{M}}_{GPC}(\vx_{GPC})$) and LPCs (i.e., $\tilde{\mathcal{M}}_{LPC}(\vx_{LPC})$). 
We can expect that using the latent variables of a DGM involves a low dimensionality, albeit at the cost of a complex input-output relationship. On the other hand, employing GPCs decomposition likely results in a considerably simpler input-output relationship but necessitates a larger number of variables, as the entire input domain needs to be approximated. In cases where the underlying forward model permits, the use of LPCs projection becomes viable. This parametrization offers straightforward input-output characteristics as the GPCs approach but with fewer parameters, thus facilitating the training of a surrogate.

To effectively combine MCMC methods with surrogate modeling, it is essential to ensure a high level of accuracy in the output domain. 
To achieve this goal, we propose decoupling the parameterizations used for inversion and surrogate modeling.
The latent representation provided by the DGM is used to evaluate $P(\vz)$ and explore the posterior. Once a prior has been defined \textit{and} a surrogate modeling strategy devised, a Metropolis-Hastings MCMC algorithm can be used to sample the posterior distribution $P(\vz|\vy)$ \citep{hastings1970monte}. A sample of the posterior in the physical space, $P(\mathcal{G}_{DGM}({\vz})|\vy)$, is then also available via mere application of the $\mathcal{G}_{DGM}$ to draws from $P(\vz|\vy)$.
For each step in the MCMC process, we consider three strategies with surrogates operating on latent variables, on GPCs and or LPCs associated with the field $\mathcal{G}_{DGM}({\vz})$. We depict these three strategies for  a travel time tomography problem in Fig. \ref{fig:DGM_DIAGRAM}. 

\section{Application to GPR crosshole travel time tomography}
In the previous section, we briefly covered the basic principles of Bayesian methods and emphasized the significance of dimensionality reduction and surrogate modeling for their implementation. Now, we integrate these concepts to tackle GPR crosshole traveltime tomography through MCMC.
Traveltime tomography refers to various imaging methods
where the propagation of acoustic, elastic or electro-magnetic waves is used to non-destructively infer media properties.
High-resolution MCMC traveltime tomography of electromagnetic ground wave velocity distribution in the shallow subsurface can be performed using ground-penetrating radar (GPR).
GPR wave-propagation is governed by the distribution of dielectric permittivity ($\epsilon$) and electric conductivity ($\sigma$) in the subsurface. The wave propagation velocity mainly depends on permittivity, which is in turn related to porosity and water content through petrophysical relationships \citep{gloaguen2005borehole}. 
GPR data can be collected in a variety of configurations, with crosshole designs being particularly well suited for groundwater investigations \citep{labrecque2002three,annan2005gpr}.

\subsection{Experimental setup}
\begin{figure}
\centering
\includegraphics[width=1\textwidth]{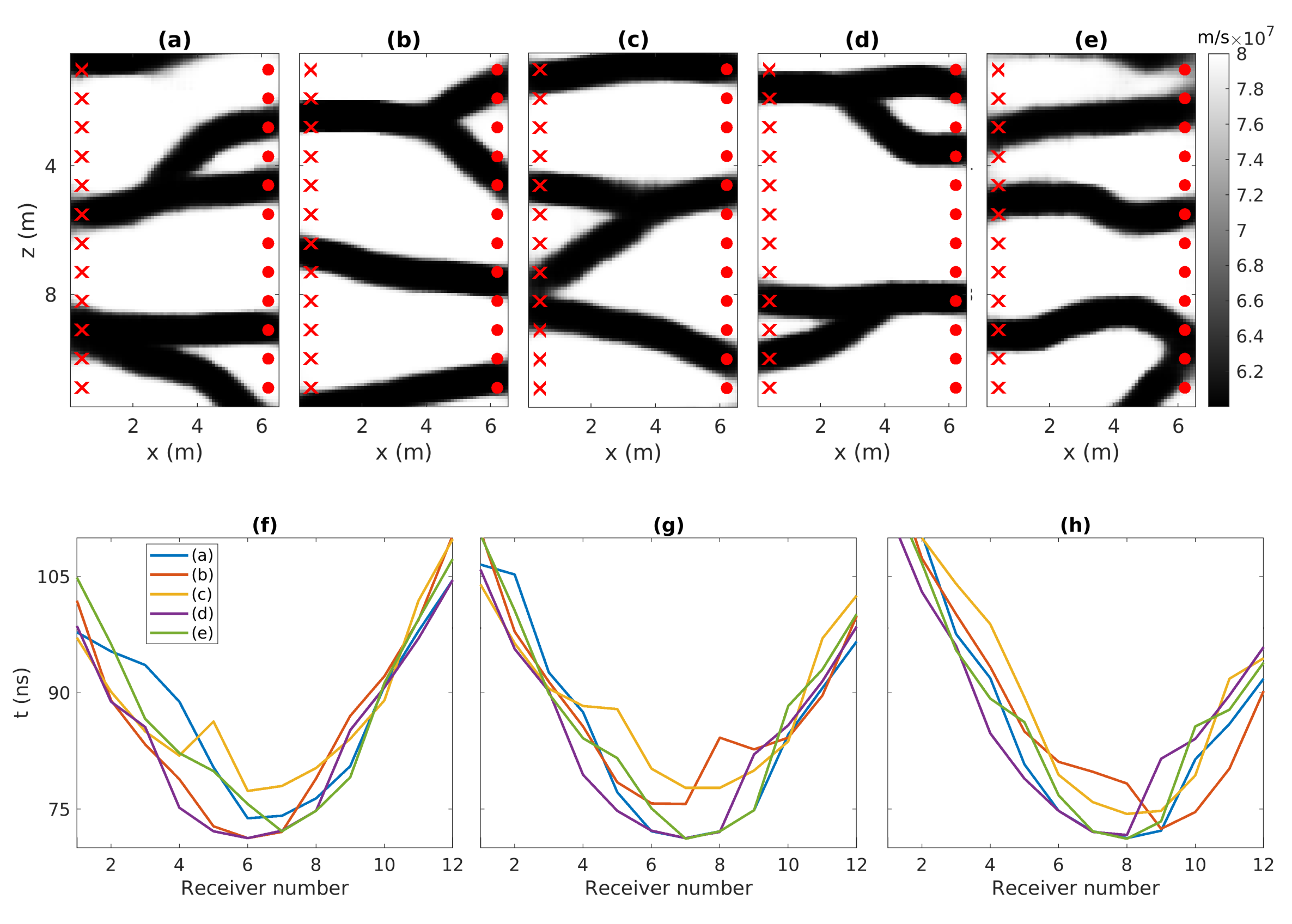}
\caption{\label{fig:001} (a-e) Representative velocity fields generated by the decoder of the employed VAE; crosses and circles stand for sources and receivers, respectively. (f-h) Corresponding exemplary traveltime gathers.}
\end{figure}

We target lossless media represented by binary images with two facies of homogeneous GPR velocity ($6\cdot 10^7$ and $8\cdot 10^7$ $m/s$) resembling river channels
\citep{strebelle2002conditional,zahner2016image,laloy2018training}. 
For the representation of the prior and posterior exploration, we consider coordinates induced by a  VAE (in the following the subscript $_{VAE}$ refers to this specific DGM parametrization)  as recent research suggests that their lower degree of nonlinearity in the corresponding networks compared with GANs makes them more amenable for modeling and inversion \citep{lopez2021deep,levy2022variational}.
The details of the VAE utilized in this study can be found in \citet{laloy2017inversion} with the dimension of the latent space being $20$.
As for the output, we consider arrival-times  associated with the crosshole configuration displayed in Figs. \ref{fig:001}(a-e) with 12 evenly spaced sources and receivers located in two vertically-oriented boreholes. The distance between the boreholes is 4.6 m, while the spacing between sources/receivers is 0.9 m, such that a total of $144$ traveltimes are collected. We employ both an eikonal and a $2D$ Finite Difference Time Domain (FDTD) solver to simulate noise-free propagation of GPR waves \citep{irving2006numerical,hansen2013sippi}. For the FDTD code each source is characterized by a $100$ MHz Blackman–Harris function, while perfectly matched layers surrounding the propagation domain are used to prevent spurious reflections from contaminating the data, while appropriate space-time grids are employed to avoid instability and dispersion artefacts. Traveltimes are picked automatically based on a threshold associated with the relative maximum amplitude of each source-receiver pair.

\label{GPR_CASE}
 
\subsection{PCE modeling within MCMC}
\label{parametrization}

For surrogate-based modeling, we rely on PCE modeling due to its efficiency, flexibility  and ease of deployment \citep{xiu2002wiener,blatman2011adaptive,luethen2021sparse,metivier2020efficient}. 
We here summarize the most relevant aspects of PCE modeling with a surrogate $\mathcal{M}_{GPC}({\vx}_{GPC})$, but identical considerations would apply to $\mathcal{M}_{DGM}$, $\mathcal{F}$ or $\mathcal{M}_{LPC}$, albeit with relevant caveats or advantages that will be discussed below.
PCE approximate functions in terms of linear combinations of orthonormal multivariate polynomials $\Psi_{\boldsymbol{\alpha}}$:
\begin{equation}
 \tilde{\mathcal{M}}_{GPC}(\vx_{GPC}) = \sum_{\boldsymbol{\alpha} \in  \mathcal{A}} a_{\boldsymbol{\alpha}} \Psi_{\boldsymbol{\alpha}}(\vx_{GPC}),
 \label{PCE_Total}
\end{equation}
where $M$ is the dimension of $\vx_{GPC}$ and  $\mathcal{A}$ is a subset of $\mathbb{N}^{{M}}$ implementing a truncation scheme to be set based on accuracy requirements and available computational resources \citep{xiu2002wiener}. 
The training of the coefficients $a_{\ve{\alpha}}$ is computationally unfeasible when the input domain is high-dimensional (the case for a surrogate $\tilde{\mathcal{F}}(\vu)$ of  $\mathcal{F}(\vu)$). Moreover, when the imposed truncation pattern cannot fully account for the degree of non-linearity of the underlying model (the case for a surrogate $\tilde{\mathcal{M}}_{VAE}(\vz)$ of $\mathcal{M}_{VAE}(\vz)$), the still unbiased PCE predictions are inevitably affected even if the input domain is low-dimensional. On the other hand, using  tailored PCA decomposition as required by $\tilde{\mathcal{M}}_{LPC}(\vx_{LPC})$ of $\mathcal{M}_{LPC}(\vx_{LPC})$ could decrease computational burden and increase accuracy.
In any case, 
the surrogate forward modeling predictor can be evaluated at a negligible cost by direct computation of Eq.~\eqref{PCE_Total} and its accuracy 
estimated using a validation set or cross-validation techniques \citep{blatman2011adaptive,UQdoc_14_104}. 

We here test the three strategies discussed in section \ref{effective}
 using PCE for surrogate modeling in conjunction with the VAE parametrization for prior-sampling. For each strategy we build a corresponding PCE to model traveltime arrivals using the Matlab Package UQlab \citep{marelli2014uqlab,UQdoc_14_104}.
To offer a fair comparison, we employ the same training and validation datasets for all proposed schemes. 

In the first strategy, referred to as VAE-PCE, the input for the PCE modeling are  $20$-dimensional $\vz$ vectors 
mapping the latent space into the physical one, that is: $\mathcal{G}_{VAE}({\vz}) = {\vu}$ \citep{lopez2021deep}.

\begin{figure}
\centering
\includegraphics[width=1\textwidth]{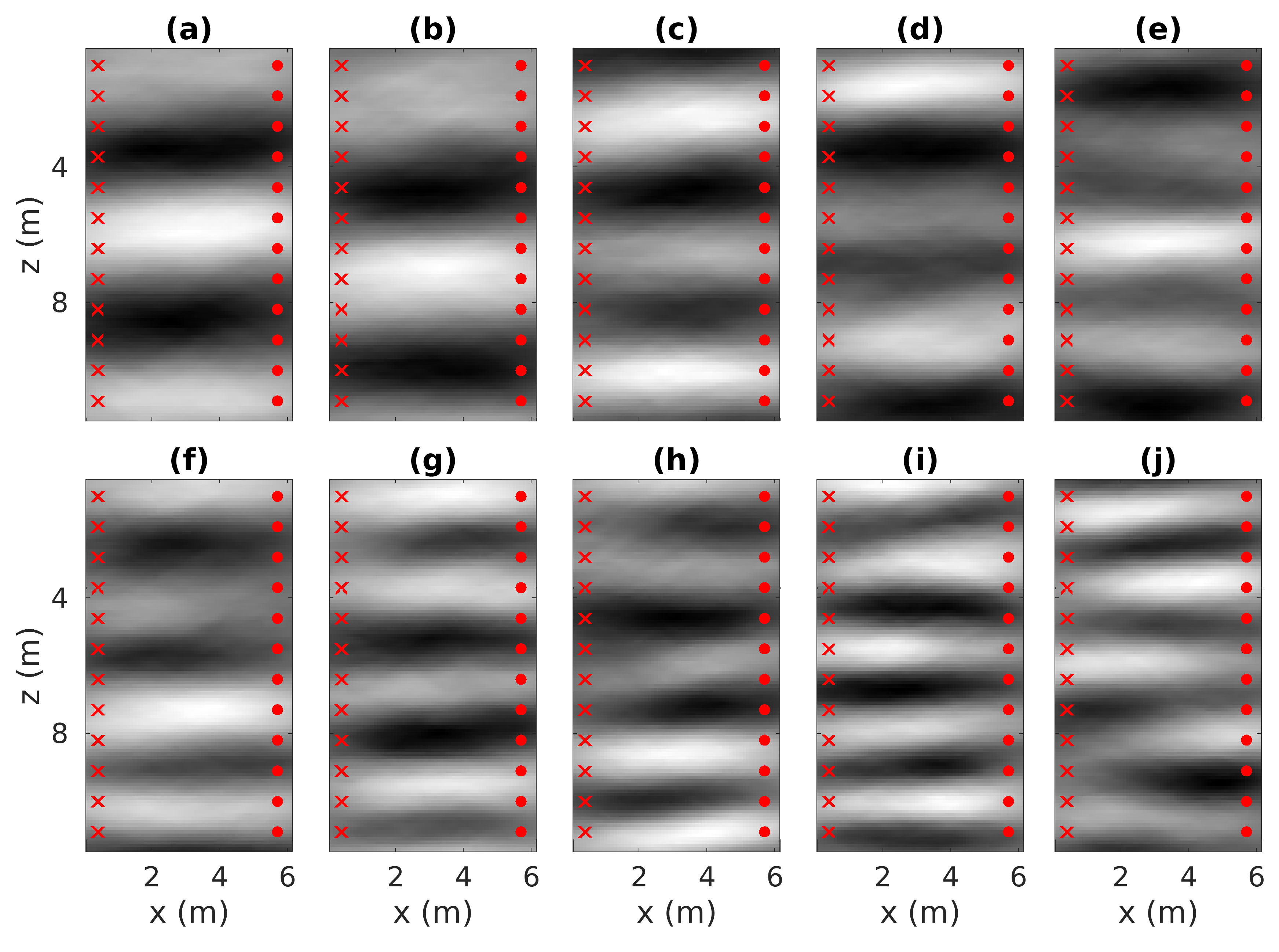}
\caption{\label{fig:002}(a-j) The first five GPCs in the input domain corresponding to entire slowness fields. Crosses and circles stand for sources and receivers, respectively.}
\end{figure}

The second strategy, in the following Global-PCA-PCE, uses a similar approach to \citet{meles2022bayesian}, with inputs of the PCE modeling defined in terms of projections on prior-informed PCA components spanning the entire  domain.
More specifically, in the Global-PCA-PCE approach we randomly create a total of $1000$ slowness realizations $\mathcal{G}_{VAE}({\vz})$ from the prior and compute the corresponding principal components (see Fig. \ref{fig:002}).
The input for PCE in the Global-PCA-PCE approach are then the projections of $\mathcal{G}_{VAE}({\vz})$ on a to-be-chosen number of such GPCs. 
Following \citet{meles2022bayesian}, all PCA processes are defined in terms of slowness..

\begin{figure}
\centering
\includegraphics[width=.9\textwidth]{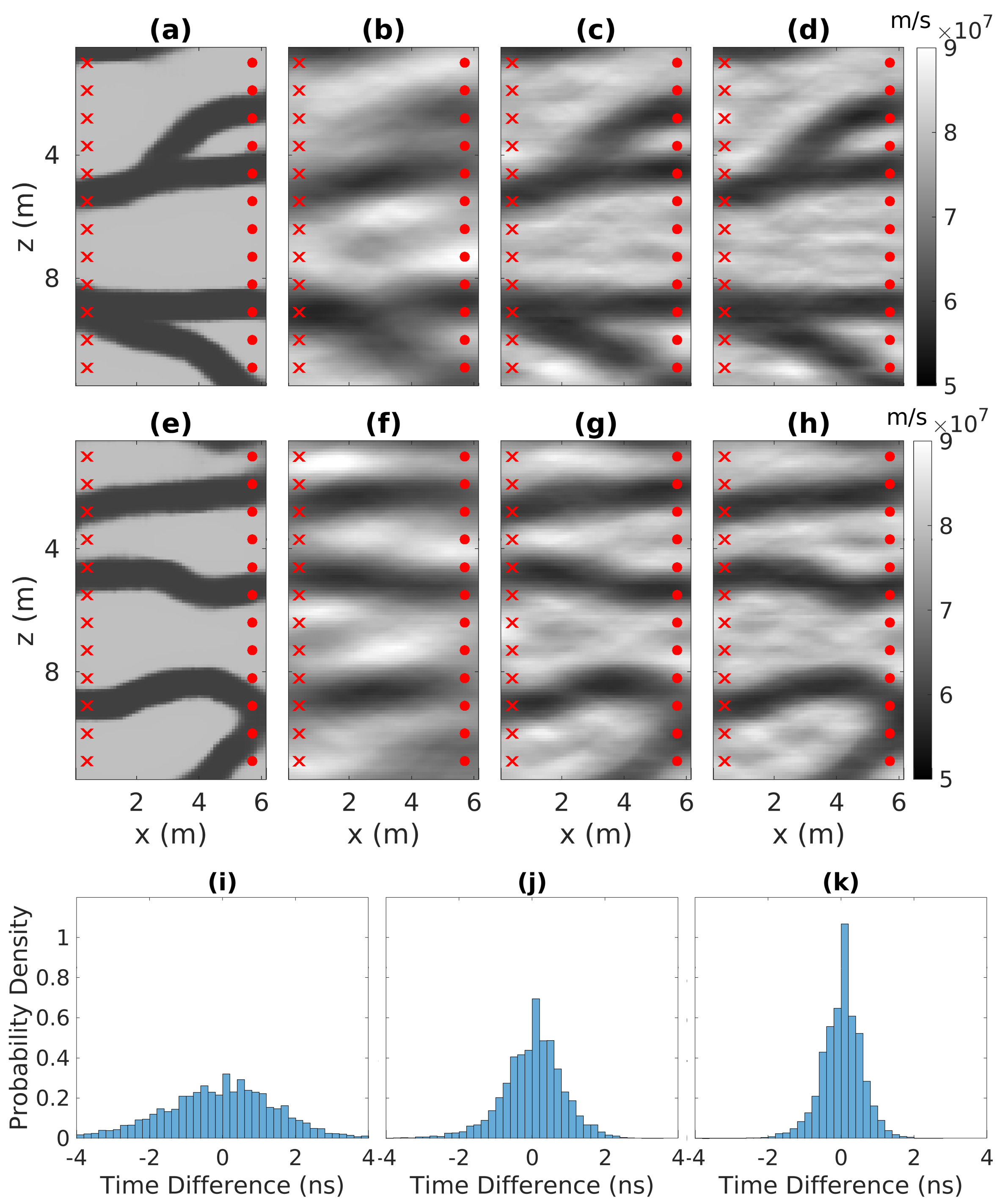}
\caption{\label{fig:003} (a) and (e) Random velocity distributions and the corresponding representations using the first (b) and (f) 30, (c) and (g) 60, and (d) and (h) 90 GPCs defined with the Global approach; (i-k) corresponding histograms of traveltime residuals based on simulations on the true field and partial reconstructions for 100 velocity distributions. Crosses and circles stand for sources and receivers, respectively.}
\end{figure}

The effective dimensionality of the input with respect to $\mathcal{M}_{GPC}$, that is, the number of GPCs representing the input, is not a-priori defined. Following a similar approach to \citet{meles2022bayesian}, the effective dimensionality is here assessed by comparison with the reference solution in the output domain with respect to the noise level.
In Figs. \ref{fig:003}(a) and (e), two velocity distributions are shown next to the approximate representations (Fig. \ref{fig:003}(b-d) and (f)-(h)) obtained by projecting them on $30, 60 $ and $90$  GPCs, respectively. As expected, the reconstruction quality improves as more principal components are included. 
To quantify the faithfulness of the various reduced parametrizations in terms of the output, we consider 100 realizations of the generative model, and compute the resulting histograms of the traveltime residuals using the reference forward solver. The root-mean-square error (in the following, rmse) of the misfit between the data associated with the original distribution and its projections on $30$, $60$ and $90$ principal components, shown in Figs. \ref{fig:003}(i)-(k), are $1.60$, $0.85$ and $0.55$ ns, respectively, that are to be compared to the expected level of GPR data noise of $1$ ns for $100$ MHz data \citep{arcone1998ground}. The number of principal components required to approximate the forward solver below the expected noise level (i.e., $90$) is larger than for the example considered by \citet{meles2022bayesian} (i.e., $50$). Building a PCE on such a large basis can be challenging in terms of computational requirements and efficiency, and could lead to poor accuracy if a small training set is employed. To address this, one approach is to either reduce the number of components, which introduces larger modeling errors, or explore alternative parameterizations that offer improved computational efficiency and accuracy. In this study, the Global-PCA-PCE approach utilizes $60$ GPCs, while an alternative strategy is considered below that is based on physical considerations.

\begin{figure}
\centering
\includegraphics[width=0.8\textwidth]{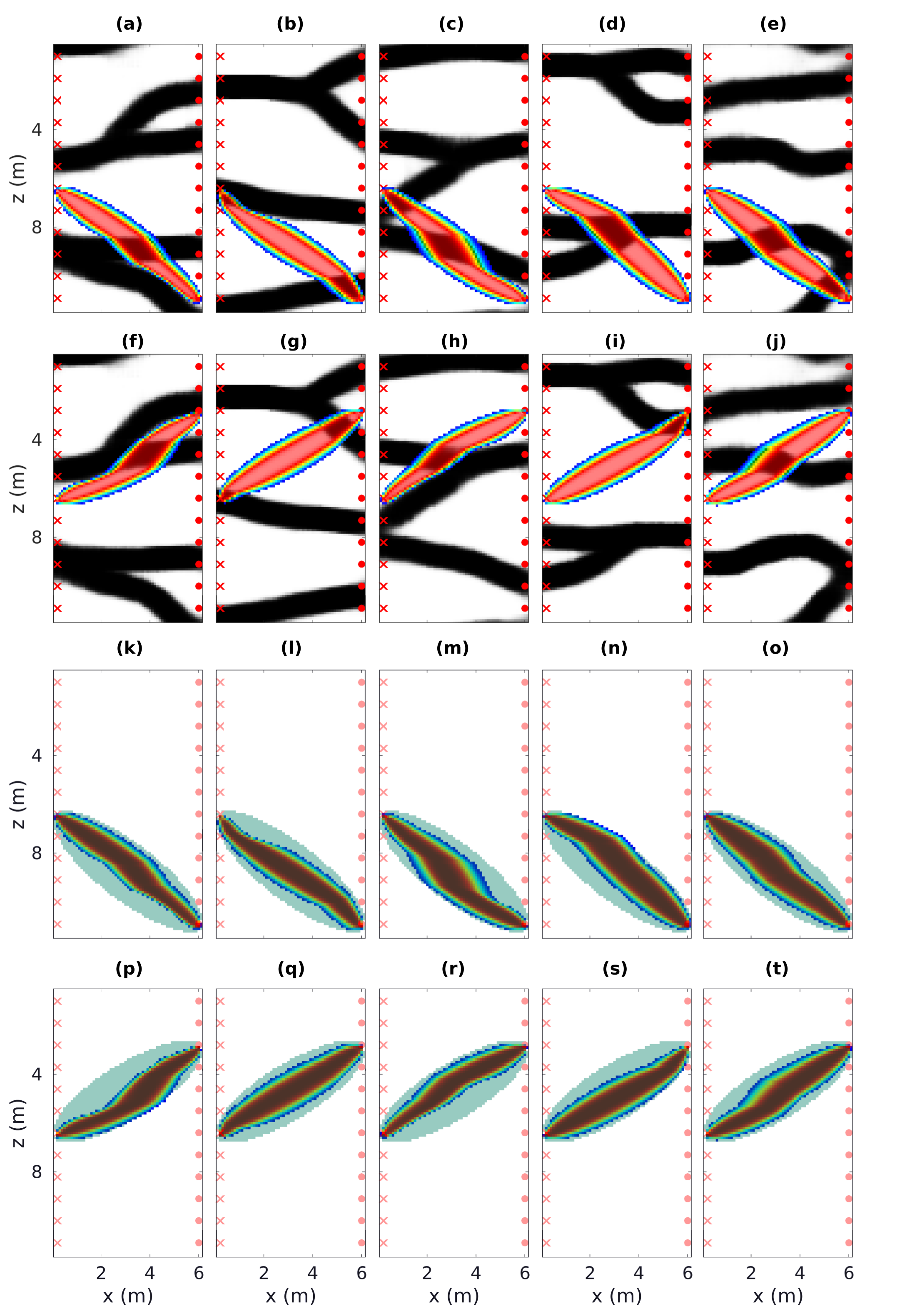}
\caption{\label{fig:004}(a)-(j) For the velocity fields in Fig. \ref{fig:001}, the sensitivity-based kernels for two arbitrarily selected source/receiver pairs are shown superimposed on the corresponding velocity distributions. (k)-(t) The same sensitivity kernels as in (a)-(j), but superimposed on the cumulative kernels (green shaded areas) used to define the support of the sensitivity-based LPCs used in the Local-PCA-PCE approach.}
\end{figure}

As mentioned in section \ref{effective}, an improved parametrization for surrogate modeling can sometimes be found by considering the forward problem's specific characteristics. The GPCs in the Global-PCA-PCE approach refer to the input field in its entirety. However, the actual first-arrival time for a given source/receiver combination depends only on a sub-domain of the entire slowness distribution. This leads us to suggest a Local approach, in the following referred to as Local-PCA-PCE. Instead of using principal components describing the entire slowness field, we aim to employ output-specific LPCs  that characterize only the sub-domains impacting the physics of the problem   \citep{jensen2000sensitivity,husen2001local}.
We then expect that fewer LPCs are needed than GPCs to achieve the desired input/output accuracy.
In practice, the construction of LPCs involves utilizing fat-ray sensitivity kernels, which capture the sensitivity of arrival-times to small perturbations in the subsurface model, thus providing valuable insights into the regions (corresponding to the $L$ subset in  \eqref{Lforwardreduced}) that have the most significant impact on the observed data.
For a given source/receiver pair, the corresponding sensitivity kernel depends on the slowness field itself and its patterns can vary significantly (see Fig. \ref{fig:004}(a)-(j)). The sought Local decomposition needs to properly represent any possible slowness field within the prior, thus it reasonable to define it based on a representative sample of input realizations. To reduce the overall PCE computational cost it is also convenient to limit the number of used output-driven decompositions. To achieve these goals, we assume that the prior model is stationary with respect to translation. Instead of focusing on each specific source/receiver pair (a total of $144$), we can then consider each source/receiver angle (a total of $23$). We then use a total of $1000$ slowness realizations $\mathcal{G}_{VAE}({\vz})$ from the prior and build the corresponding fat-ray sensitivity kernels using the reference eikonal solver for each of the $23$ possible angles  \citep{hansen2013sippi}.
For any given angle, 
we consider a cumulative kernel consisting of the sum of the absolute values of each kernel (green areas in Fig. \ref{fig:004}(k)-(t)). Such a cumulative kernel cover an area larger than each individual kernel but is still considerably smaller than the entire slowness field. For any possible additional input model, the corresponding sensitivity kernel is then very likely to be \textit{geometrically} included in the area covered by the cumulative kernel (see Fig. \ref{fig:004}(k)-(t)). Based on this insight, we define principal components spanning only the area covered by such cumulative kernels or relevant parts thereof (e.g., a threshold can be taken into consideration to reduce the size of these sub-domains). For the practical definition of the components, the cumulative kernels are either set to $0$ or $1$ depending on whether the corresponding value is below or larger than the threshold, respectively. We then multiply point by point the slowness distributions with the cumulative kernels, and consider the principal components of such products.

\begin{figure}
\centering
\includegraphics[width=1\textwidth]{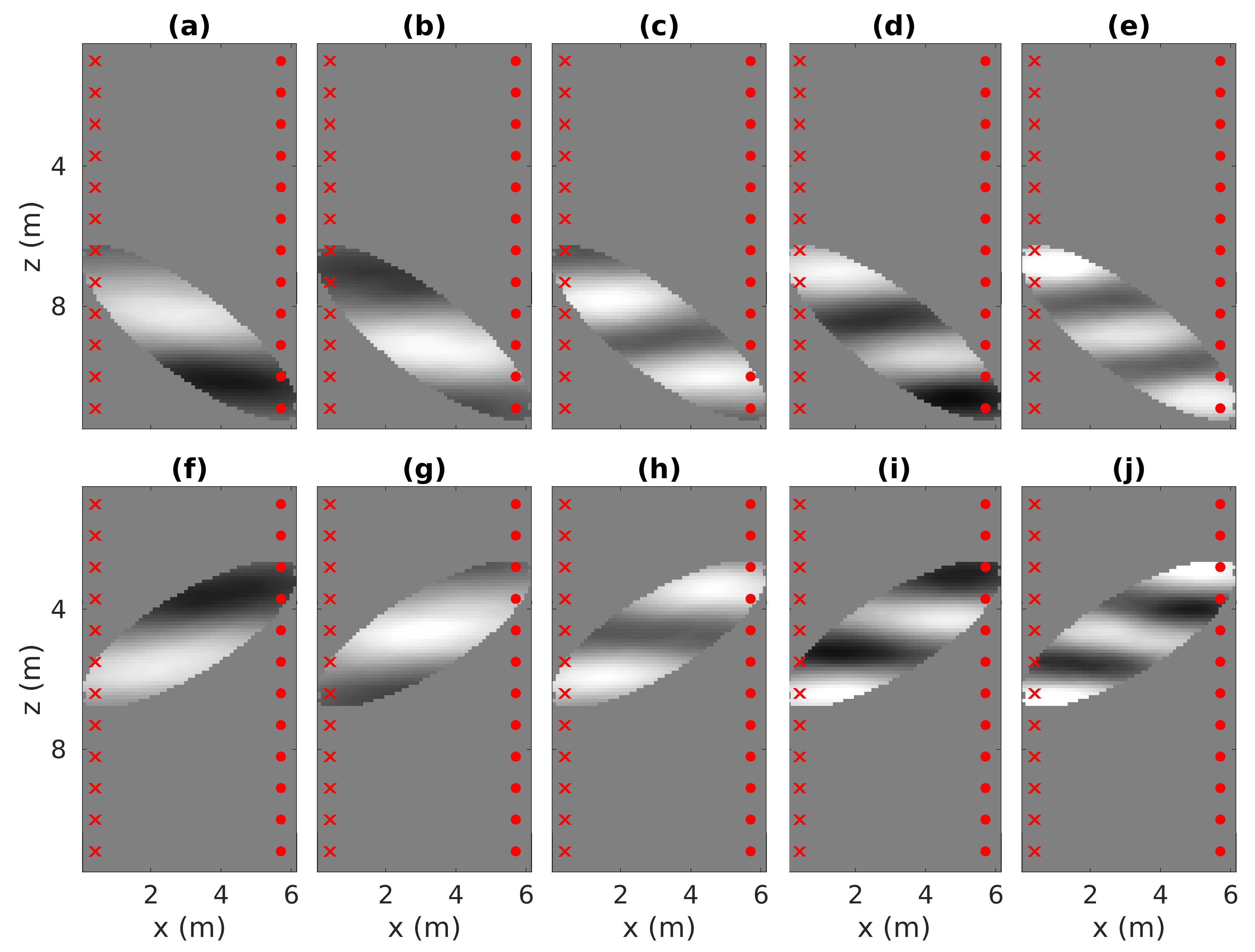}
\caption{\label{fig:005a}The first five LPCs in the input domain corresponding to the cumulative kernels associated with the source/receiver pairs considered in Figs. \ref{fig:004}(k)-(t).}
\end{figure}

The first five LPCs are shown for given source/receiver pairs (Figs. \ref{fig:005a}(a)-(e) and (f)-(j)). Note that the pattern variability is confined within the cumulative kernels,  while in the complementary areas the values are $0$. Note also that compared to the five principal components in Fig. \ref{fig:002}, higher resolution details can be identified. Given the same number of principal components, we can then expect the input to be better presented in the physically relevant sub-domain when the Local-PCA-PCE rather then the Global-PCA-PCE approach is followed.
For all source/receiver pairs corresponding to the same altitude angle, the same kernels and principal components are used, provided they are shifted to cover the appropriate geometry.

\begin{figure}
\centering
\includegraphics[width=1\textwidth]{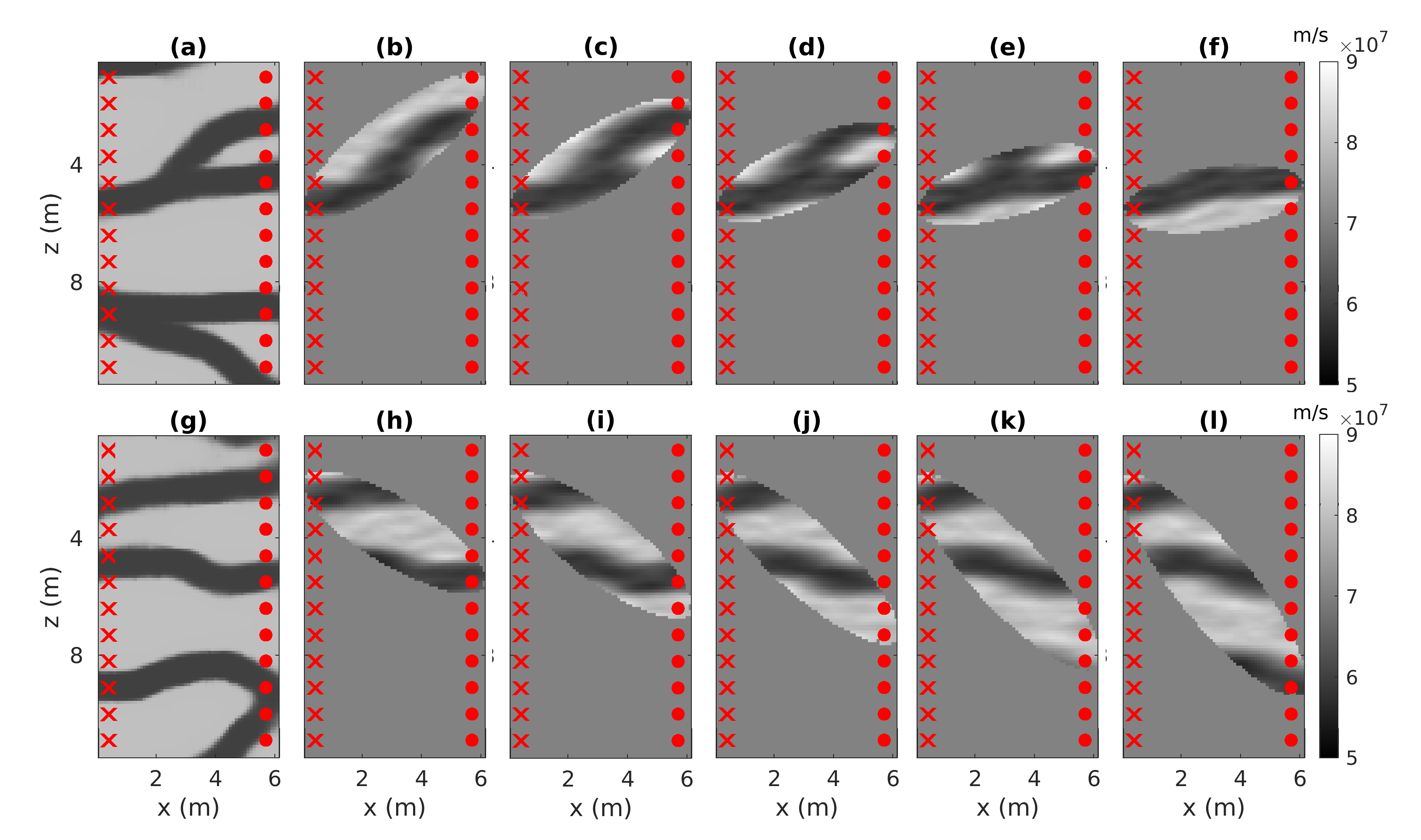}
\caption{\label{fig:005b}(a) and (g) Velocity fields with (b)-(f) and (h)-(l) the corresponding representations used for surrogate modeling based on the first 30 LPCs used in the Local-PCA-PCE approach. Different kernels are used for each source/receiver angle.}
\end{figure}

In Figs. \ref{fig:005b}(a)-(g) the two slowness distributions from Fig. \ref{fig:003} are shown next to the representations obtained by projecting them on $30$ LPCs. In the areas complementary to the sensitivity kernels, the speed is set to $0.07$ m/ns.
Input reconstructions are remarkably improved with respect to when using the same number of GPCs (compare Figs. \ref{fig:005b}(a)-(g) and (h)-(l) to Fig. \ref{fig:003}(b) and (f)) of the entire slowness field. More importantly, the modeling errors provided by using just $30$ sensitivity-based LPCs is lower than what was previously provided by $90$ standard components (i.e., rms $\approx 0.45$ ns).
By incorporating these tailored LPCs, we can attain enhanced output fidelity when utilizing truncated representations of the input. This enhanced fidelity proves particularly advantageous for the implementation of PCE, allowing for more precise and efficient modeling. Consequently, this approach holds substantial promise in achieving superior accuracy and computational efficiency in PCE-based analyses. 

In summary, we have introduced three different parametrizations to be used as input for PCE. We consider coordinates inherited by the VAE, and principal components derived by considering either entire slowness fields or sensitivity-based sub-domains. We refer to these three parametrizations in what follows as VAE-PCE, Global-PCA-PCE and Local-PCA-PCE, respectively.

\subsection{PCE performance}
\label{PCEperformance}
We here analyze the PCE performance of the different input parametrizations for surrogate modeling introduced in Section \ref{parametrization}. In agreement with \citet{meles2022bayesian}, we consider for each surrogate a total of $900$ training sets and a polynomial degree $p$ of five when applied on eikonal data.

\begin{figure}
\centering
\includegraphics[width=1\textwidth]{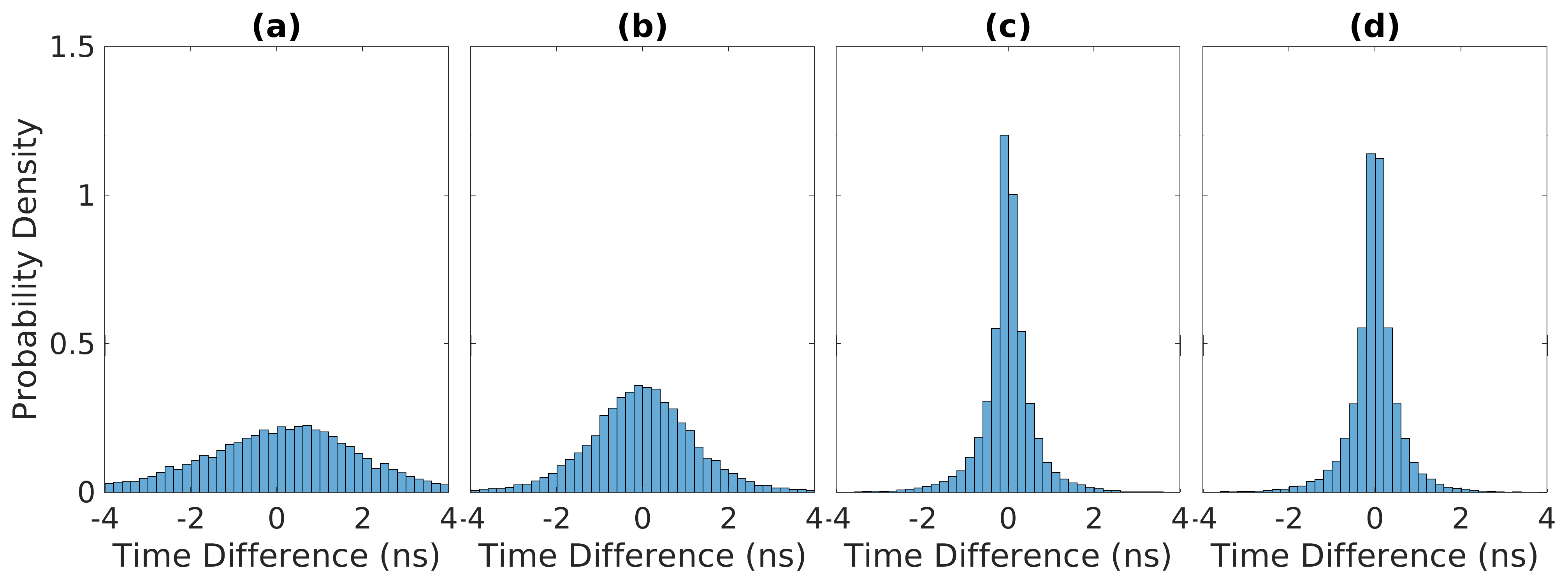}
\caption{\label{fig:006a} (a)-(c): Histograms of the model error with respect to the PCE prediction when using the VAE (20 input parameters), Global (60 input parameters) and Local (30 input parameters per angle) parameterizations of the input in the PCE-based surrogate modeling, respectively, using the eikonal solver to compute the training set. (d) Histogram of the model error using Local parameters and FDTD reference data.} 
\end{figure}

When the VAE parametrization is used as input, the PCE performance is rather poor, with a rmse of $2.01$ ns, which is well beyond the expected noise level and is consequently considered unsatisfactory (Fig. \ref{fig:006a}(a)).
The poor performance is due to the highly non-linear map $\mathcal{M}_{VAE}$. In such a scenario,  PCE  does not provide accurate results even if the physical input of the validation set is exactly defined by  $\mathcal{G}_{VAE}(\vz)=\vu$. In this scheme, note that the evaluation of $\mathcal{G}_{VAE}(\vz)$ is not required to compute the corresponding PCE. 

\begin{figure}
\centering
\includegraphics[width=1\textwidth]{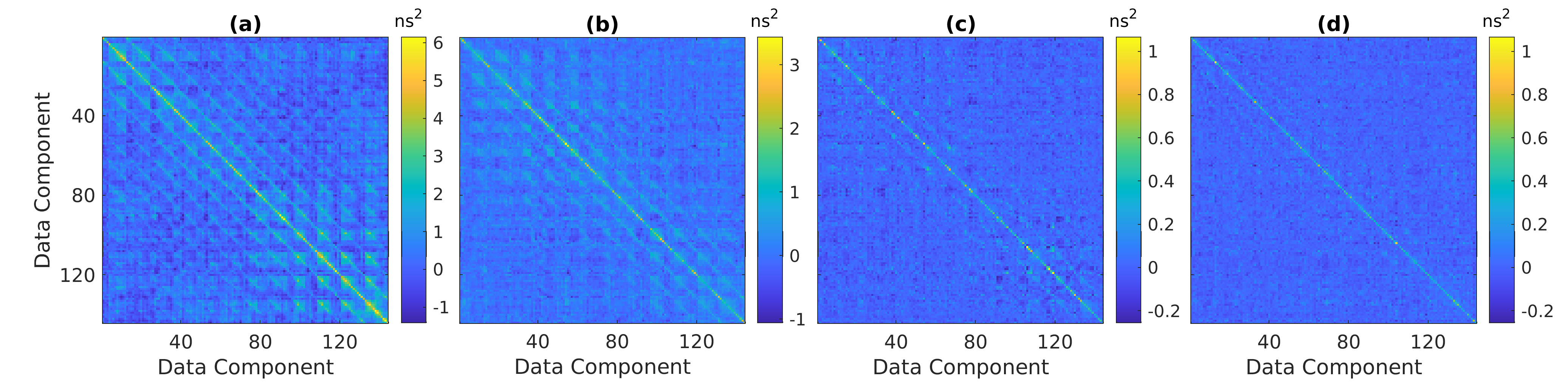}
\caption{\label{fig:006b} Model error covariance matrices associated with PCE-based surrogate modeling based on: (a) VAE-PCE, (b) Global-PCA-PCE and (c) Local-PCA-PCE schemes trained with 900 eikonal datasets.
(d) Model error covariance matrix for the Local-PCA-PCE scheme trained with FDTD data.}
\end{figure}

Despite the comparatively poor reconstruction of the input (e.g., $\mathcal{G}_{G/LPC}(\vx) \approx \vu$) provided by the PCA approaches, the corresponding parametrizations perform well when used as input to build PCE surrogates, with the Global-PCA-PCE approach being outperformed by the Local-PCA-PCE scheme in terms of accuracy (rmse of $1.31$ and $0.68$ ns, respectively, see Figs. \ref{fig:006a}(b) and (c) for the corresponding histograms). In both of these cases, the PCE operates on more variables (i.e., $60$ and $30$  for the Global-PCA-PCE and Local-PCA-PCE parametrizations, respectively, versus $20$ for the VAE-PCE scheme).
Moreover, the input for the Global-PCA-PCE and Local-PCA-PCE schemes are projections of images, which require the evaluation of $\mathcal{G}_{VAE}(\vz)$, on G/LPCs. As such, evaluation of the corresponding PCEs is computationally more expensive for the PCA-based approaches than for the VAE-PCE case.
In the Local-PCA-PCE approach, for each of the $23$ angles considered, training involves randomly chosen source/receiver- pair data associated with identical angles, while the final rmse is computed on the standard $144$ traveltime gathers.
For the Local-PCA-PCE scheme we also consider training and validation using FDTD data in addition to the eikonal data discussed above. Results are similar and well below the noise, with an rmse of $0.65$ ns (the corresponding histogram is displayed in Fig. \ref{fig:006a}(d)). All PCE results are unbiased and the model errors have Gaussian-like distributions.
Depending on the input parameterization, PCEs approximate eikonal and FDTD solvers to different degrees. Figure \ref{fig:006b} represent the corresponding covariance matrices accounting for the modeling error of each surrogate model.

We now discuss the computational burden of each strategy when running on a workstation equipped with 16GB DDR4 of RAM and powered by a 3.5GHz Quad-Core processor running Matlab and Python on Linux. We emphasize that our goal with the present manuscript is to propose novel methods to conjugate VAE and PCE rather than offer optimized implementations. 

There are up to three relevant computational steps in the execution of a forward run for the VAE-PCE, Global-PCA-PCE, Local-PCA-PCE, eikonal and FDTD simulations, namely the evaluation of the physical input, $\mathcal{G}_{VAE}({\vz})$, its decomposition on either GPCs or LPCs and the actual evaluation of the forward or surrogate model. Not all methods require each of these steps. The VAE-PCE model, for example, is not a function  of $\mathcal{G}_{VAE}({\vz})$ but depends on ${\vz}$ only. Evaluation of the VAE-PCE model is extremely fast both when involving one or 35 (as in the MCMC inversion discussed below that is based on 35 chains) simultaneous model runs, taking on average $\approx 0.06s$  and $\approx 0.08s$, respectively.
Evaluation of the Python-based decoder $\mathcal{G}_{VAE}({\vz})$ required for all forward models except the VAE-PCE, is the bottleneck of the Matlab algorithms used herein, requiring $\approx 1.35s$  and $\approx 1.43s$, respectively, when operating on one or 35 input when considering the eikonal solver. Nevertheless, this cost could be reduced by either implementing the decoder and the PCE within the same Python environment or by optimizing the calls to the Matlab/Python scripts in our code for greater efficiency.    
When in its native environment, evaluation of $\mathcal{G}_{VAE}({\vz})$ is actually very fast, taking only $\approx 0.005$  and $\approx 0.08s$ when operating on one or 35 inputs, respectively. Still, note that this cost is overall negligible even in our non-optimized setting when considering expensive physics-based forward solvers such as FDTD.
Only the Global-PCA-PCE and Local-PCA-PCE strategies require PCA decompositions. 
The Global-PCA-PCE approach is faster, requiring only up to $\approx 0.002s$ and $\approx 0.05s$ when applied to one and 35 input elements, respectively, while the Local-PCA-PCE method is slower, taking up to $\approx 0.06s$ and  $\approx 0.23s$ in the same situation.
For the Global-PCA-PCE method the cost of a single forward run is $\approx 0.52s$, which is significantly more than for VAE-PCE. The difference between these two PCE strategies can be attributed to the significantly larger number of input parameters of the Global-PCA-PCE with respect to the VAE-PCE scheme (i.e., $60$ vs $20$). Note that the PCE model evaluations are vectorized and, therefore, the cost is almost the same when applied to $35$ input ($\approx 0.57s$).
Moreover, the computational cost of the Global-PCA-PCE method could be reduced by applying a PCA decomposition of the output, akin to what is proposed in \citet{meles2022bayesian}.
Despite involving fewer variables than the Global-PCA-PCE approach, the Local-PCA-PCE method is slightly more computationally demanding with a cost of $\approx 0.64s$  and $\approx 0.65s$, respectively, when operating on one or $35$ input, respectively.
The increase in cost compared to the Global-PCA-PCE method depends on the fact that for each source-receiver angle ($\theta$), the Local-PCA-PCE utilizes its own polynomial basis, denoted as  $\Psi^{\theta}$ in the following.

In comparison with the PCE methods, the cost of the reference eikonal solver is basically a linear function of the number of input distributions it operates on. A single run requires $\approx 0.05s$, while for $35$ velocity distributions the cost increases up to $\approx 1.67s$.
As such, its cost is either significantly smaller or slightly larger than what is required by the Global/Local-PCA-PCE approaches. Finally, the cost required by the reference FDTD code is $\approx 120s$ and $\approx 4200s$ if operating on either one or $35$ velocity distributions, which is orders of magnitude longer than for the eikonal or PCE models. These results are summarized in Table \ref{table:0}, where we estimate the performance of an ideally-optimized Local-PCA-PCE method  benefitting from (a) evaluating $\mathcal{G}_{VAE}({\vz})$ in its native environment and (b) using a single family of polynomials $\Psi^{\theta}$ for all angles. In numerical results not presented herein, we find that choosing \textit{any one} of the $\Psi^{\theta}$ families for all models provides nearly identical fidelity to what is achieved by using specifically tailored polynomials for each angle at the considerably smaller cost of $\approx 0.06$ and $0.16s$ when applied to either one or 35 input, respectively. While such a result cannot be generalized, it is always possible to test the corresponding PCEs accuracy with a representative evaluation set. The option of relying on a single family of polynomials for the Local-PCA-PCE method is certainly to be taken into account when optimising the approach.
\begin{table}
\begin{center}
\caption{Summary of the computational cost of the various steps for a single realization/batch of 35 input of the proposed algorithms. In addition to the strategies used in the MCMC examples discussed in the manuscript and summarized in the first four columns (i.e., the VAE-PCE, Global- and Local-PCA-PCE and eikonal schemes), we also consider the cost of the FDTD approach using the reference code (fifth column) and an optimized Local-PCA-PCE approach ideally benefitting from executing the VAE decoder in the same environment as the PCE model and based on a single polynomial family for all angles (sixth column).} 
\begin{adjustbox}{width=1\textwidth}
\begin{tabular}{||c c c c c c c||} 
 \hline
 $\mbox{}$  & VAE-PCE   & Global-PCA-PCE  & Local-PCA-PCE  & Eikonal & FDTD & Optimized Local-PCA-PCE   \\ [0.5ex] 
 \hline
 $\mathcal{G}_{VAE}({\vz})$  & 0 & $\approx 1.35/1.43 s$   & $\approx 1.35/1.43 s$  & $\approx 1.35/1.43 s$  & $\approx 1.35/1.43 s$  & $\approx 0.005/0.08 s$  \\ 
  \hline
 PCA & 0 & $\approx 0.002/0.05 s$ & $\approx 0.06/0.23 s$ & 0 & 0 & $\approx 0.06/0.23 s$ \\
 \hline
 Forward & $\approx 0.06/0.08 s$ & $\approx 0.52/0.57 s$ & $\approx 0.64/0.65 s$ & $\approx 0.05/1.67 s$ & $\approx 120/4200 s$ & $\approx 0.06/0.16 s$  \\
 \hline
 
\end{tabular}
\end{adjustbox}
\label{table:0}
\end{center}
\end{table}

\subsection{Inversion results} 

\begin{figure}
\centering
\includegraphics[width=.5\textwidth]{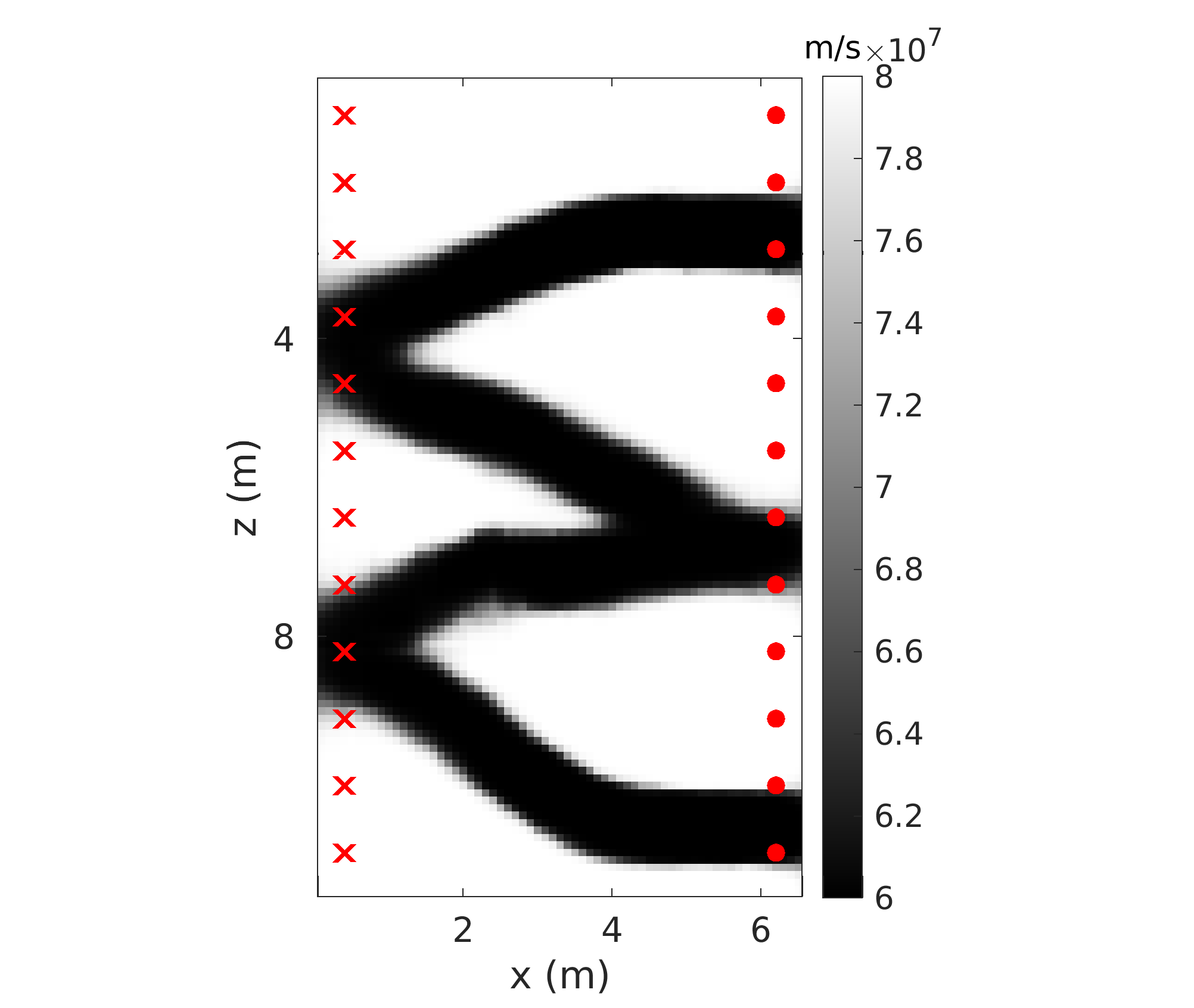}
\caption{\label{fig:007a} The reference velocity distribution used in the numerical inversion experiments.}
\end{figure}

We now explore the performance of the different input parametrizations used for PCE-based surrogate modeling, namely VAE-PCE, Global-PCA-PCE and Local-PCA-PCE, when performing probabilistic MCMC inversion. 
The inversions were carried out using the \textsc{UQLab} Matlab-based framework \citep{marelli2014uqlab,UQdoc_14_113}.
As reference model, We consider the field shown in Fig. \ref{fig:007a}, which is used to generate a total of $144$ traveltimes using the reference eikonal and FDTD solvers. Note that this field is not used to train the PCEs. Uncorrelated Gaussian noise characterized by $\sigma^2=1 ns^2$ was added to the data used for inversion. 

We use a Metropolis-Hastings algorithm, and run $35$ non-interacting Markov chains in parallel for $4\times10^5$ iterations per chain. During burn-in determined according to the Geweke method, the scaling factor of the proposal distribution is tuned such that an acceptance rate of about $30\%$ is achieved for each experiment \citep{geweke1992evaluating,brunetti2019hydrogeological}.
Finally, outlier chains with respect to the Inter Quartile-Range statistics discussed in \citet{vrugt2009accelerating} are considered
aberrant trajectories and are ignored in the analysis of the results.

\begin{figure}
\centering
\includegraphics[width=1\textwidth]{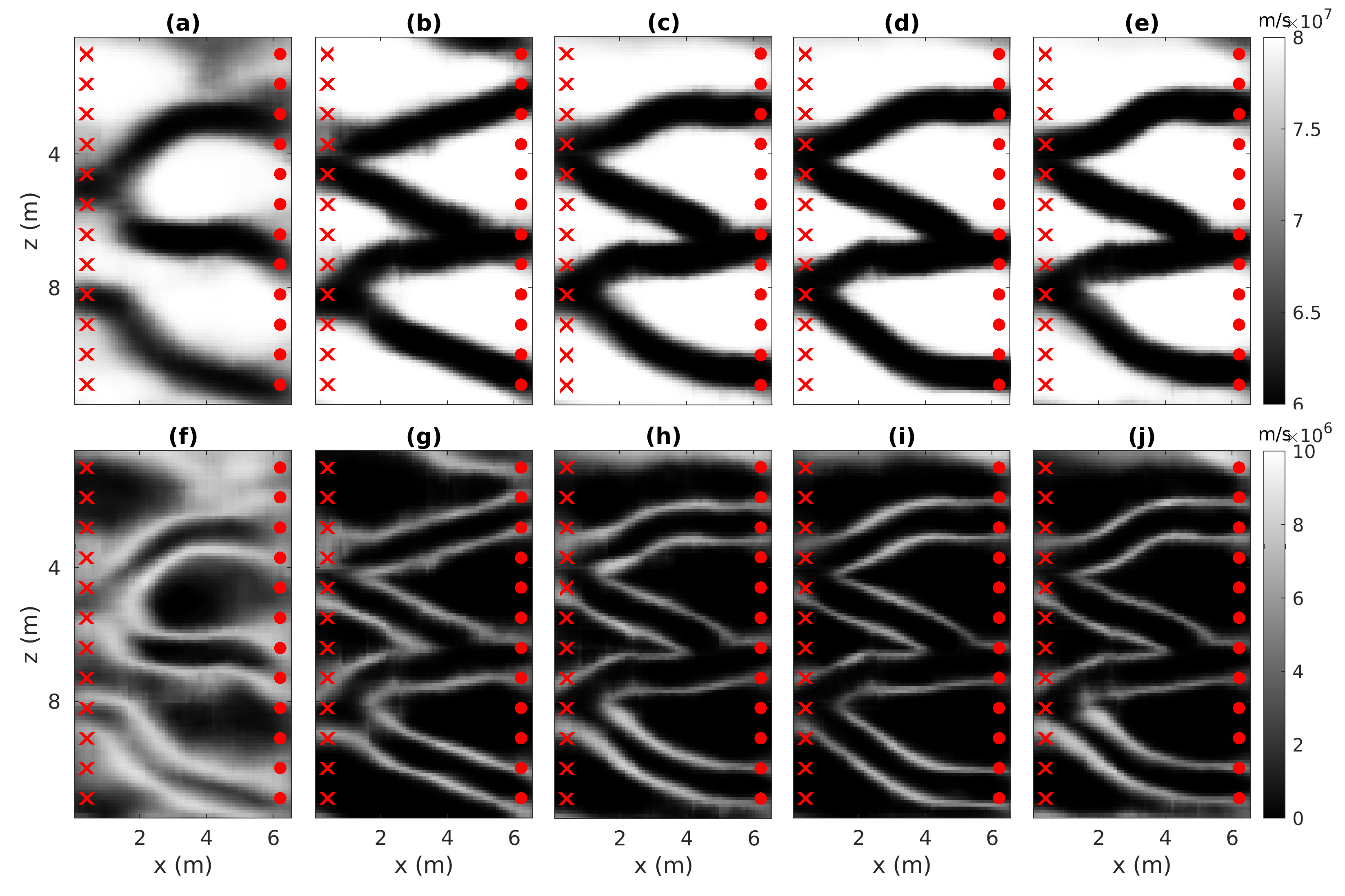}
\caption{\label{fig:007} Posterior mean and standard deviation respectively for the (a,f) VAE-PCE inversion scheme, (b, g) Global-PCA-PCE, (c,h) Local-PCA-PCE, (d,i) eikonal and (e, j) FDTD Local-PCA-PCE inversion strategies.}
\end{figure}

We first present the results for training data generated by an eikonal solver. We compare VAE-PCE, Global-PCA-PCE and Local-PCA-PCE inversion results to those achieved by employing the eikonal solver, which represent the reference solution since the full physics solver is used in the entire MCMC process.
Inversion results in terms of mean and standard deviations incorporating the model error (i.e., using the PCE-derived $\boldsymbol{C}_{Tapp}$
in Eq. \eqref{likely})
 are shown in Figs. \ref{fig:007}(a-e).
The mean of the posterior obtained using the VAE-PCE poorly resembles the reference velocity field, with relevant topological differences between the two (compare Fig. \ref{fig:007a} to Fig. \ref{fig:007}(a)). Note that the data misfit is particularly large (i.e., $3.49$ ns). This poor performance is also obtained when considering other test models (see Appendix A).
The mean of the posterior provided by the Global-PCA-PCE approach shares many features with the reference distribution, but differences are apparent in the lower and upper channelized structures.
The similarity between the posterior mean and the true distribution increases significantly when the Local-PCA-PCE is used (compare Fig. \ref{fig:007a} to Fig. \ref{fig:007}(c)). These results also show close proximity with the posterior mean solution obtained by the eikonal solver (see Fig. \ref{fig:007}(d)), that is, without any surrogate modeling.
Also, when the FDTD Local-PCA-PCE is employed, an almost identical solution to what is achieved using the Local-PCA-PCE is obtained (see Fig. \ref{fig:007}(e)).
For this alternative data set, the use of the FDTD solver in the inversion algorithm would be extremely expensive and is not considered feasible to run for comparison purposes. 
The quality of the solution offered by the surrogate on FDTD data can be heuristically appreciated by noting its similarity to results obtained by the Local-PCA-PCE based on eikonal data, which in turn produces results close to those of the eikonal solver on eikonal data. 
Although not strictly consequential, it is to be expected that the results offered by the surrogate-based on FD data would also be similar to those that would have been obtained if using the FDTD solver on FDTD data. 

\begin{figure}
\centering
\includegraphics[width=1\textwidth]{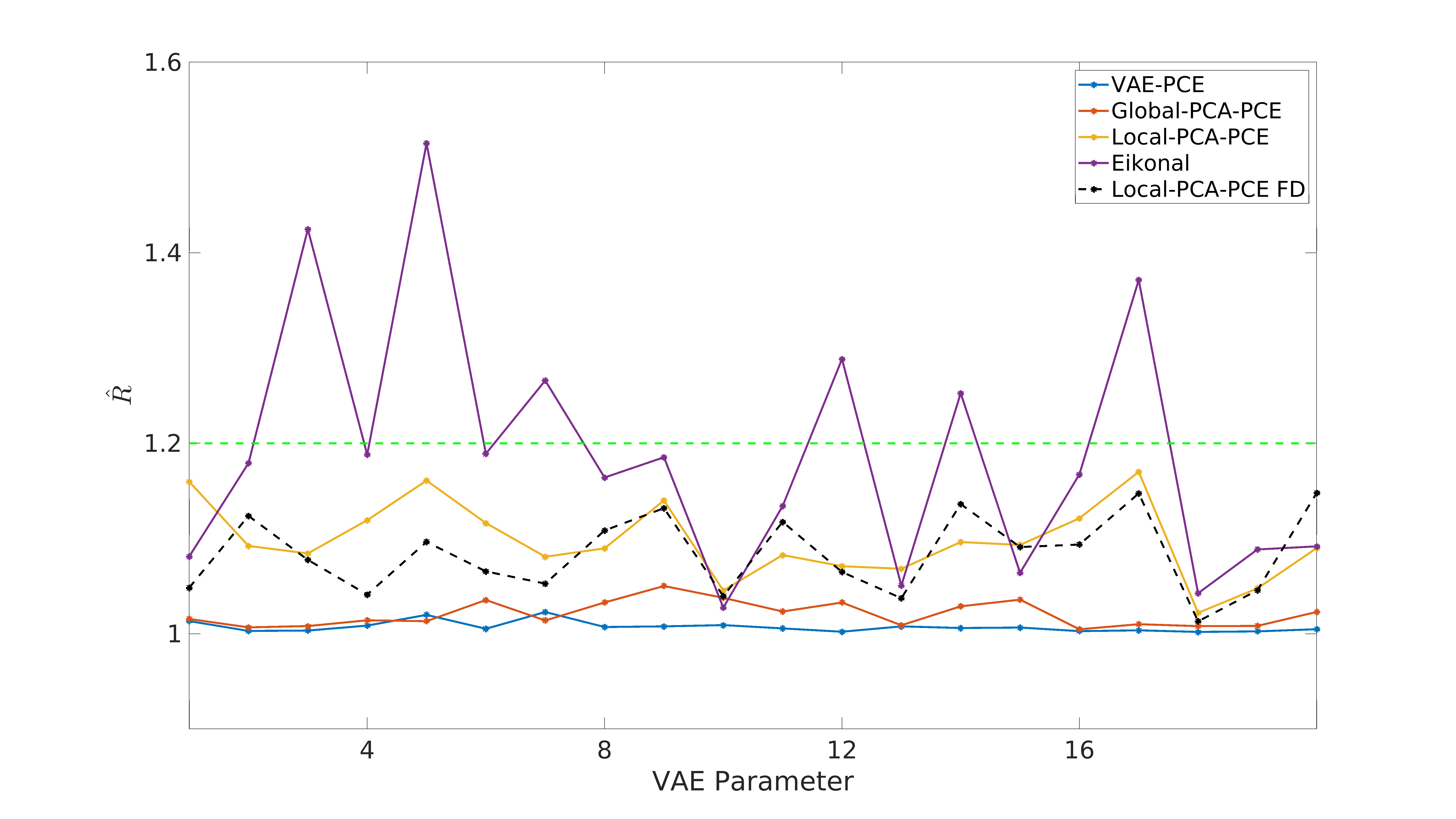}
\caption{\label{fig:009} Gelman-Rubin statistics for the various inversion strategies using 35 chains after $4\times10^5$ iterations per chain.}
\end{figure}

High standard deviations of the posterior distribution are distributed in wide domains of the image when VAE-PCE is used (see Fig. \ref{fig:007}f). In contrast, when Global-PCA-PCE (Fig. \ref{fig:007}g), Local-PCA-PCE (Fig. \ref{fig:007}h) and eikonal (Fig. \ref{fig:007}i) solvers are used, high standard deviation values are found mainly along channel boundaries in agreement with other studies \citep{zahner2016image}
Convergence is assessed using the potential-scale reduction factor $\hat{R}$ that compares the variance of the individual Markov chains with the variance of all the Markov chains merged together \citep{gelman1992inference} calculated from the second half of the chains. Convergence is usually assumed if $R<1.2$ for all parameters. In our experiments, full convergence for all of the $20$ parameters is achieved when the VAE-PCE, the Global-PCA-PCE and the Local-PCA-PCE approaches are used. Six parameters do not converge when the eikonal solver is employed, but the values $R$ are nevertheless close to 1.2 (see Fig. \ref{fig:009}). 
Further quantitative assessments can be achieved by comparing the reference input and the corresponding inversion solutions in terms of input domain  root-mean-square error (in the following, RMSE), structural similarity (in the following, SSIM) and rmse in the output domain \citep{gneiting2007strictly,levy2022using}.
Among these metrics, SSIM specifically evaluates the structural similarity between images, emphasizing their underlying patterns and details. It assigns a value between -1 and 1, with -1 indicating a notable dissimilarity and 1 denoting perfect similarity.
Again, we see that the VAE-PCE performs poorly, with a low SSIM  value. Better results are provided by the Global-PCA-PCE strategy. The Local-PCA-PCE scheme results are the closest to the reference solutions achieved using the eikonal solver. It is found that the FDTD Local-PCA-PCE performs similarly to the Local-PCA-PCE strategy.
The results for all considered metrics are summarized in Table \ref{table:2}.

\begin{table}
\begin{center}
\caption{Assessment of the inversion results in the input and output domains for the various surrogate modeling strategies.}
\begin{tabular}{||c c c c||} 
 \hline
 Model & RMSE Mean Velocity  & SSIM Mean Velocity &   rmse  Mean Output \\ [0.5ex] 
 \hline
 VAE-PCE & $8.01\cdot 10^6$ m/s & 0.30 & 3.49 ns  \\ 
 \hline
 Global-PCA-PCE & $5.38\cdot 10^6$ m/s & 0.54 & 1.49 ns  \\
 \hline
 Local-PCA-PCE & $2.67\cdot 10^6$ m/s & 0.73 & 1.15 ns   \\
 \hline
 Eikonal & $1.57\cdot 10^6$ m/s & 0.87 & 1.01 ns   \\ 
 \hline
 FD Local-PCA-PCE & $3.06\cdot 10^6$ m/s & 0.71 & 1.15 ns  \\ 
  \hline
\end{tabular}
\label{table:2}
\end{center}
\end{table}

\begin{figure}
\centering
\includegraphics[width=1\textwidth]{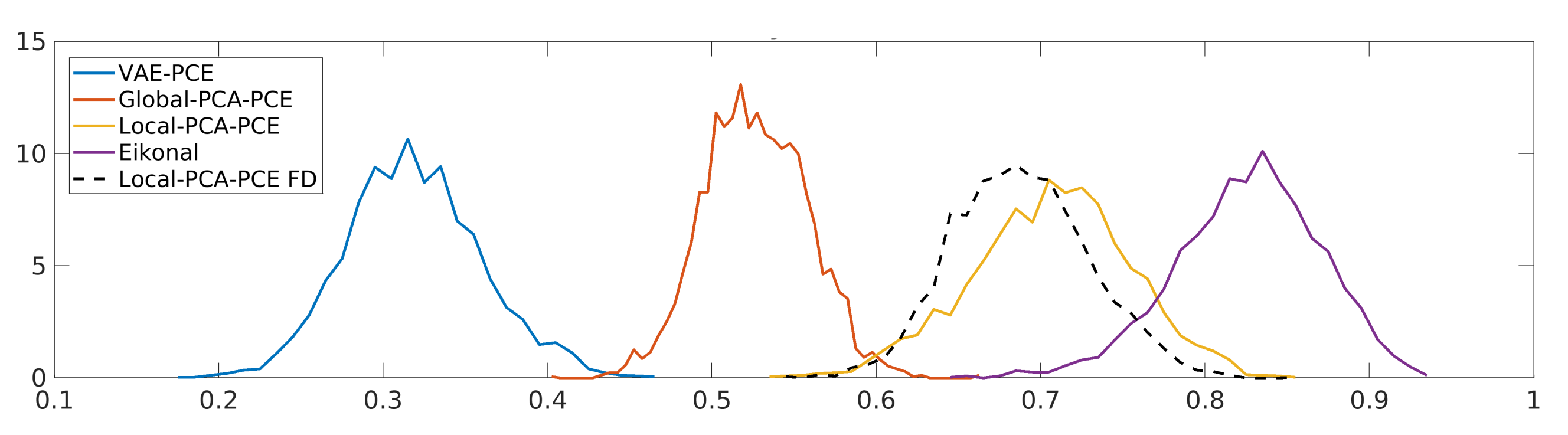}
\caption{\label{fig:012} Histograms of SSIM values across the posterior distributions for the various inversion strategies.}
\end{figure}

\begin{figure}
\centering
\includegraphics[width=1\textwidth]{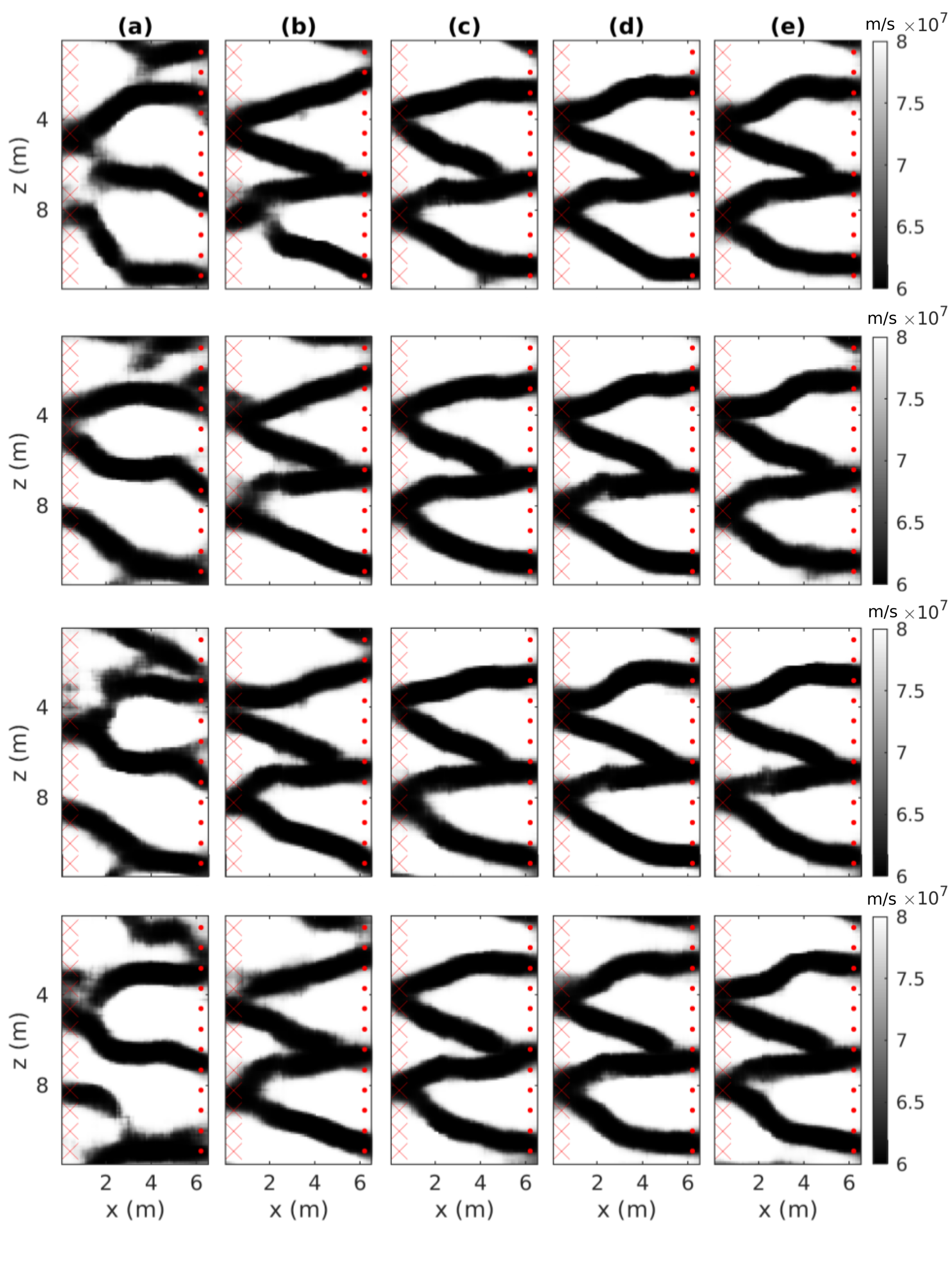}
\caption{\label{fig:013} Random samples from the posterior obtained by employing the (a) VAE-PCE, (b) Global-PCA-PCE, (c) Local-PCA-PCE, (d) eikonal and (e) FD Local-PCA-PCE strategies. The results in c-e are visually similar.}
\end{figure}

We also consider histograms of SSIM values in the corresponding posterior distributions (Fig. \ref{fig:012}). The VAE-PCE posterior has low similarity with the reference model, with the maximum SSIM value being below 0.5. Closer proximity is found among samples obtained using the Global-PCA-PCE approach, a trend that is further improved when considering the Local-PCA-PCE scheme that shows some overlap with the results of the reference eikonal inversion. Note again that the statistics of the FDTD Local-PCA-PCE algorithm is again similar to the Local-PCA-PCE scheme.
These findings can be further appreciated by considering random posterior realizations for each of the inversion strategies (see Fig. \ref{fig:013}).

\section{Discussion}
Deep generative models offer a flexible framework for encoding complex spatial priors, enabling the description of intricate input distributions. However, proper characterization of prior distributions alone does not ensure efficient estimation of posterior distributions when using MCMC methods. In many practical situations, the use of surrogate models becomes beneficial or even essential for evaluating the likelihood functions effectively.
Surrogate modeling with PCE has become widespread in many disciplines. The massive decrease of the computational costs associated with PCE is achieved by approximating demanding computational forward models with simple and easy-to-evaluate functions. A key requirement is that the number of input variables describing the computational model is relatively small (i.e., up to a few tens) and that the target model can be approximated by truncated series of low-degree multivariate polynomials.
The number of coefficients defining the PCE model grows polynomially in both the size of the input and the maximum degree of the expansion. When the reference full-physics model response is highly nonlinear in its input parameters, the problem is typically non-tractable \citep{torre2019data}.
Since the high-fidelity mapping of complex prior distributions provided by DGMs is based on highly non-linear relationships between latent variables and physical domains, replicating the performance of such networks and/or composite functions thereof (e.g., $\mathcal{M}_{VAE}=\mathcal{F}\circ G_{VAE}$ in Eq. \ref{forwardcoordinates_01}) using PCE is problematic.
To circumvent this challenge, we have explored two PCA-based decompositions that facilitated the straightforward implementation of PCE. One decomposition was designed to encompass the entire input domain, while the other specifically focused on subdomains of particular physical relevance. While this latter concept is investigated here in the context of traveltime tomography, the integration of problem-related PCs operating synergistically with DGM-defined latent parametrizations has the potential for broader applications.

Whatever the choice of input coordinates, for example, based on PCA or local properties of the input, the determining criterion for evaluating the quality of the corresponding PCE should always be performance on a representative validation set.
In case of PCA, the lower bound of prediction misfit rmse can be a priori estimated by comparing the accuracy of the reference model acting on the full and the compressed domains, that is,  ${\mathcal{M}_{G/LPC}}(\vx_{full})$ and $ \mathcal{M}_{G/LPC}(\vx)$. In our case, such lower bounds for the Global-PCA-PCE approach operating with $60$ components  is $0.85$ ns. Using the Local-PCA-PCE scheme with $30$ components only, the rmse drops to $0.55$ ns.
However, the accuracy of the corresponding Global/Local-PCA-PCE is worse, with rmse of $1.31$ and $0.67$ ns, respectively, mainly due to the small size of the training set. Note that while the lower bound of PCA decreases when more PCs are taken into account, the corresponding accuracy of PCE is limited by the size of the training set. Increasing the number of components can actually worsen PCE performance if the training set is not adequate to determine the polynomial coefficients, which, as mentioned, grow significantly with input size. In our case, using $90$ components would imply an rmse of $1.39$ ns, which is worse than what was obtained by the $60$ components PCE. Note that while for the VAE-PCE method the parametrization is inherited by the DGM architecture, for the Global/Local-PCA-PCE schemes the number of variables/PCs has to be chosen based on accuracy and computational burden analysis, with the computational cost increasing with the size of the input. In this study using $30$ components for both the Global-PCA-PCE and the Local-PCA-PCE strategies would have reduced the accuracy of the Global-PCA-PCE method. On the other hand, using $60$ components for the Local-PCA-PCE scheme would have decreased its computational efficiency.

Both the VAE-PCE and Global-PCA-PCE consist of $144$ (i.e., the total number of source/receiver combinations) different PCE models operating, for each traveltime simulation, on an identical input, that is, the latent variables of the DGM or the $60$ PCs characterizing the entire physical domain.
On the other hand, the Local-PCA-PCE scheme consists of $23$ (i.e., the total number of source/receiver altitude angles) different models operating, for each traveltime simulation, on $144$ different input, that is, the $30$ LPCs characterizing each local sub-domain.
Since each of the 23 models operates on specific LPCs, the corresponding families of orthonormal polynomials $\Psi^{\theta}$ are different. This is in contrast with the Global-PCA-PCE method, for which each model operates via a single family of polynomials.
Thus, the Local-PCA-PCE scheme is computationally more demanding than the Global-PCA-PCE (see Table \ref{table:0}). However, the use of a single family of polynomials can also be considered for the Local-PCA-PCE method, resulting in shorter run time. 
When considering computational performance, an optimal implementation of $\mathcal{G}_{VAE}({\vz})$ should also be sought. 

In this study, to determine the minimum number of G/LPCs for constructing an accurate PCE, we assess the lower bound of output prediction misfit rmse as a function of the number of G/LPCs used. We project the input onto subsets of G/LPCs, typically ranging in the tens. This process generates non-binary images, which are then utilized to compute the output using the reference forward modeling. 
Alternatively, we could consider re-binarizing the reconstructed images as done in \citet{thibaut2021new}. This approach would bring the projected input back into the prior, but this property is not necessarily relevant for the determination of PCE accuracy. Irrespective of the chosen reconstruction algorithm, the Local approach maintains a significant advantage over the Global method. When using an equal number of components, LPCs, in fact, consistently yield superior approximations of the relevant input compared to GPCs.

We have seen that once a DGM-based latent parametrization has been found to reduce the effective dimensionality of the input domain and, based on PCA decompositions, high fidelity PCEs have been trained, MCMC inversion can be efficient.
Relying on PCE rather than advanced deep learning methods for surrogate modeling 
can be advantageous in terms of ease of implementation, as potentially complex training of a neural network is not needed.
Many effective sampling methods, such as Adaptive Metropolis, Hamiltonian Monte Carlo, or the Affine Invariant Ensemble Algorithm \citep{duane1987hybrid,haario2001adaptive,goodman2010ensemble} could be considered in our workflow instead of the current use of a standard Metropolis-Hastings sampling algorithm.

Adaptation of the Global-PCA-PCE strategy presented here could easily be employed for other imaging problems such as active or passive seismic tomography at different scales \citep{bodin2009seismic,galetti2017transdimensional}. On the other hand, implementation of the Local-PCA-PCE schemes would depend on the properties of the corresponding sensitivity kernels, which would require more careful evaluation and problem-specific design.
\section{Conclusions}
Low-dimensional latent variables associated with deep generative models, such as VAE, optimally conform to complex priors, and provide an ideal basis to explore posterior model distributions using MCMC strategies. 
MCMC methods can also benefit greatly from surrogate modeling based on PCE, provided the forward model can be approximated by low-degree multivariate polynomials.
However, this type of latent variable models tend to have a highly non-linear relation to data, and are, thus, poorly approximated by low-degree PCEs. As such, performing PCE-accelerated MCMC inversion based on a latent parametrization for both inversion and as imput to surrogate modeling leads to large posterior uncertainty due to the need to account for important modeling errors in the likelihood function.
In the context of GPR traveltime tomography, PCE-based surrogate modeling using VAE latent variables as input results in modeling errors that are well beyond the noise level.
By separating the parametrization used for inversion and the one used as input for surrogate modeling, we can circumvent this problem and perform MCMC in a latent space defined by deep generative models while surrogate modeling is approximated by PCEs operating on globally or locally-defined principal components.
We find that these two approaches largely outperform surrogate modeling based on VAE input parametrizations. 
For the channelized structures considered, modeling errors are comparable to the typical observational errors when PCE are based on globally-defined principal components and significantly lower when locally defined principal components are considered. Generally speaking, using PCE significantly reduces the computational burden of MCMC, but it can be successfully employed to perform non-linear MCMC inversion only if the corresponding modeling error is not excessively large.
In this manuscript we have shown how PCE based on VAE parametrizations performs poorly in MCMC inversion, whereas PCE based on globally and locally-defined principal components produce results comparable or close to those obtained using full-physics forward solvers.
The methods presented herein are extendable to other problems involving wave-based physics of similar complexity. 
\section{Acknowledgments}
We acknowledge the feedback from associate editor Prof. Andrew Valentine (Durham University) and reviewer Dr. Robin Thibaut (Ghent University). Niklas Linde, Macarena Amaya and Shiran Levy acknowledge support by the Swiss National Science Foundation (project number: 184574).  
\section{Data availability}
The data underlying this article are available upon request to the corresponding author.

\bibliographystyle{apalike}
\bibliography{paper}

\providecommand{\noopsort}[1]{}\providecommand{\singleletter}[1]{#1}%
\begin{thebibliography}{}

\bibitem[Annan, 2005]{annan2005gpr}
Annan, A.~P. (2005).
\newblock {GPR} methods for hydrogeological studies.
\newblock In {\em Hydrogeophysics}, pages 185--213. Springer.

\bibitem[Arcone et~al., 1998]{arcone1998ground}
Arcone, S.~A., Lawson, D.~E., Delaney, A.~J., Strasser, J.~C., and Strasser,
  J.~D. (1998).
\newblock Ground-penetrating radar reflection profiling of groundwater and
  bedrock in an area of discontinuous permafrost.
\newblock {\em Geophysics}, 63(5):1573--1584.

\bibitem[Aster et~al., 2018]{aster2018parameter}
Aster, R.~C., Borchers, B., and Thurber, C.~H. (2018).
\newblock {\em Parameter {E}stimation and {I}nverse {P}roblems}.
\newblock Elsevier.

\bibitem[Blatman and Sudret, 2011]{blatman2011adaptive}
Blatman, G. and Sudret, B. (2011).
\newblock Adaptive sparse polynomial chaos expansion based on least angle
  regression.
\newblock {\em Journal of Computational Physics}, 230(6):2345--2367.

\bibitem[Bodin and Sambridge, 2009]{bodin2009seismic}
Bodin, T. and Sambridge, M. (2009).
\newblock Seismic tomography with the reversible jump algorithm.
\newblock {\em Geophysical Journal International}, 178(3):1411--1436.

\bibitem[Boutsidis et~al., 2008]{boutsidis2008unsupervised}
Boutsidis, C., Mahoney, M.~W., and Drineas, P. (2008).
\newblock Unsupervised feature selection for principal components analysis.
\newblock In {\em Proceedings of the 14th ACM SIGKDD international conference
  on Knowledge discovery and data mining}, pages 61--69.

\bibitem[Brunetti et~al., 2019]{brunetti2019hydrogeological}
Brunetti, C., Bianchi, M., Pirot, G., and Linde, N. (2019).
\newblock Hydrogeological model selection among complex spatial priors.
\newblock {\em Water Resources Research}, 55(8):6729--6753.

\bibitem[Chipman et~al., 2001]{chipman2001practical}
Chipman, H., George, E.~I., McCulloch, R.~E., Clyde, M., Foster, D.~P., and
  Stine, R.~A. (2001).
\newblock The practical implementation of {B}ayesian model selection.
\newblock {\em Lecture Notes-Monograph Series}, pages 65--134.

\bibitem[Duane et~al., 1987]{duane1987hybrid}
Duane, S., Kennedy, A.~D., Pendleton, B.~J., and Roweth, D. (1987).
\newblock Hybrid {M}onte {C}arlo.
\newblock {\em Physics Letters {B}}, 195(2):216--222.

\bibitem[Galetti et~al., 2017]{galetti2017transdimensional}
Galetti, E., Curtis, A., Baptie, B., Jenkins, D., and Nicolson, H. (2017).
\newblock Transdimensional {L}ove-wave tomography of the {B}ritish {I}sles and
  shear-velocity structure of the {E}ast {I}rish {S}ea {B}asin from
  ambient-noise interferometry.
\newblock {\em Geophysical Journal International}, 208(1):36--58.

\bibitem[Gelman and Rubin, 1992]{gelman1992inference}
Gelman, A. and Rubin, D.~B. (1992).
\newblock Inference from iterative simulation using multiple sequences.
\newblock {\em Statistical {S}cience}, 7(4):457--472.

\bibitem[Geweke, 1992]{geweke1992evaluating}
Geweke, J. (1992).
\newblock Evaluating the accuracy of sampling-based approaches to the
  calculations of posterior moments.
\newblock {\em Bayesian {S}tatistics}, 4:641--649.

\bibitem[Giannakis et~al., 2021]{giannakis2021fractal}
Giannakis, I., Giannopoulos, A., Warren, C., and Sofroniou, A. (2021).
\newblock Fractal-constrained crosshole/borehole-to-surface full-waveform
  inversion for hydrogeological applications using ground-penetrating radar.
\newblock {\em IEEE Transactions on Geoscience and Remote Sensing}.

\bibitem[Gloaguen et~al., 2005]{gloaguen2005borehole}
Gloaguen, E., Marcotte, D., Chouteau, M., and Perroud, H. (2005).
\newblock Borehole radar velocity inversion using cokriging and cosimulation.
\newblock {\em Journal of Applied Geophysics}, 57(4):242--259.

\bibitem[Gneiting and Raftery, 2007]{gneiting2007strictly}
Gneiting, T. and Raftery, A.~E. (2007).
\newblock Strictly proper scoring rules, prediction, and estimation.
\newblock {\em Journal of the American Statistical Association},
  102(477):359--378.

\bibitem[Goodfellow et~al., 2020]{goodfellow2020generative}
Goodfellow, I., Pouget-Abadie, J., Mirza, M., Xu, B., Warde-Farley, D., Ozair,
  S., Courville, A., and Bengio, Y. (2020).
\newblock Generative adversarial networks.
\newblock {\em Communications of the ACM}, 63(11):139--144.

\bibitem[Goodman and Weare, 2010]{goodman2010ensemble}
Goodman, J. and Weare, J. (2010).
\newblock Ensemble samplers with affine invariance.
\newblock {\em Communications in Applied Mathematics and Computational
  Science}, 5(1):65--80.

\bibitem[Haario et~al., 2001]{haario2001adaptive}
Haario, H., Saksman, E., and Tamminen, J. (2001).
\newblock An adaptive {M}etropolis algorithm.
\newblock {\em Bernoulli}, pages 223--242.

\bibitem[Hansen et~al., 2014]{hansen2014accounting}
Hansen, T.~M., Cordua, K.~S., Jacobsen, B.~H., and Mosegaard, K. (2014).
\newblock Accounting for imperfect forward modeling in geophysical inverse
  problems—exemplified for crosshole tomography.
\newblock {\em Geophysics}, 79(3):H1--H21.

\bibitem[Hansen et~al., 2013]{hansen2013sippi}
Hansen, T.~M., Cordua, K.~S., Looms, M.~C., and Mosegaard, K. (2013).
\newblock Sippi: A matlab toolbox for sampling the solution to inverse problems
  with complex prior information: Part 2—application to crosshole {GPR}
  tomography.
\newblock {\em Computers \& Geosciences}, 52:481--492.

\bibitem[Hastings, 1970]{hastings1970monte}
Hastings, W.~K. (1970).
\newblock Monte carlo sampling methods using markov chains and their
  applications.
\newblock {\em Biometrika}, 57(1):97--109.

\bibitem[Higdon et~al., 2015]{higdon2015bayesian}
Higdon, D., McDonnell, J.~D., Schunck, N., Sarich, J., and Wild, S.~M. (2015).
\newblock A {B}ayesian approach for parameter estimation and prediction using a
  computationally intensive model.
\newblock {\em Journal of Physics G: Nuclear and Particle Physics},
  42(3):034009.

\bibitem[Ho et~al., 2020]{ho2020denoising}
Ho, J., Jain, A., and Abbeel, P. (2020).
\newblock Denoising diffusion probabilistic models.
\newblock {\em Advances in neural information processing systems},
  33:6840--6851.

\bibitem[Husen and Kissling, 2001]{husen2001local}
Husen, S. and Kissling, E. (2001).
\newblock Local earthquake tomography between rays and waves: fat ray
  tomography.
\newblock {\em Physics of the Earth and Planetary Interiors},
  123(2-4):127--147.

\bibitem[Irving and Knight, 2006]{irving2006numerical}
Irving, J. and Knight, R. (2006).
\newblock Numerical modeling of ground-penetrating radar in 2-{D} using
  {MATLAB}.
\newblock {\em Computers \& Geosciences}, 32(9):1247--1258.

\bibitem[Jensen et~al., 2000]{jensen2000sensitivity}
Jensen, J.~M., Jacobsen, B.~H., and Christensen-Dalsgaard, J. (2000).
\newblock Sensitivity kernels for time-distance inversion.
\newblock {\em Solar Physics}, 192(1):231--239.

\bibitem[Jetchev et~al., 2016]{jetchev2016texture}
Jetchev, N., Bergmann, U., and Vollgraf, R. (2016).
\newblock Texture synthesis with spatial generative adversarial networks.
\newblock {\em arXiv preprint arXiv:1611.08207}.

\bibitem[Jolliffe and Cadima, 2016]{jolliffe2016principal}
Jolliffe, I.~T. and Cadima, J. (2016).
\newblock Principal component analysis: a review and recent developments.
\newblock {\em Philosophical Transactions of the Royal Society A: Mathematical,
  Physical and Engineering Sciences}, 374(2065):20150202.

\bibitem[Kingma and Welling, 2013]{kingma2013auto}
Kingma, D.~P. and Welling, M. (2013).
\newblock Auto-encoding variational bayes.
\newblock {\em arXiv preprint arXiv:1312.6114}.

\bibitem[LaBrecque et~al., 2002]{labrecque2002three}
LaBrecque, D., Alumbaugh, D.~L., Yang, X., Paprocki, L., and Brainard, J.
  (2002).
\newblock Three-dimensional monitoring of vadose zone infiltration using
  electrical resistivity tomography and cross-borehole ground-penetrating
  radar.
\newblock In {\em Methods in Geochemistry and Geophysics}, volume~35, pages
  259--272. Elsevier.

\bibitem[Laloy et~al., 2018]{laloy2018training}
Laloy, E., H{\'e}rault, R., Jacques, D., and Linde, N. (2018).
\newblock Training-image based geostatistical inversion using a spatial
  generative adversarial neural network.
\newblock {\em Water Resources Research}, 54(1):381--406.

\bibitem[Laloy et~al., 2017]{laloy2017inversion}
Laloy, E., H{\'e}rault, R., Lee, J., Jacques, D., and Linde, N. (2017).
\newblock Inversion using a new low-dimensional representation of complex
  binary geological media based on a deep neural network.
\newblock {\em Advances in Water Resources}, 110:387--405.

\bibitem[Lataniotis et~al., 2020]{lataniotis2020extending}
Lataniotis, C., Marelli, S., and Sudret, B. (2020).
\newblock Extending classical surrogate modeling to high dimensions through
  supervised dimensionality reduction: a data-driven approach.
\newblock {\em International Journal for Uncertainty Quantification}, 10(1).

\bibitem[Levy et~al., 2022a]{levy2022using}
Levy, S., Hunziker, J., Laloy, E., Irving, J., and Linde, N. (2022a).
\newblock Using deep generative neural networks to account for model errors in
  markov chain monte carlo inversion.
\newblock {\em Geophysical Journal International}, 228(2):1098--1118.

\bibitem[Levy et~al., 2022b]{levy2022variational}
Levy, S., Laloy, E., and Linde, N. (2022b).
\newblock Variational bayesian inference with complex geostatistical priors
  using inverse autoregressive flows.
\newblock {\em Computers \& Geosciences}, page 105263.

\bibitem[Lopez-Alvis et~al., 2021]{lopez2021deep}
Lopez-Alvis, J., Laloy, E., Nguyen, F., and Hermans, T. (2021).
\newblock Deep generative models in inversion: The impact of the generator's
  nonlinearity and development of a new approach based on a variational
  autoencoder.
\newblock {\em Computers \& Geosciences}, 152:104762.

\bibitem[L\"uthen et~al., 2021]{luethen2021sparse}
L\"uthen, N., Marelli, S., and Sudret, B. (2021).
\newblock Sparse polynomial chaos expansions: Literature survey and benchmark.
\newblock {\em SIAM/ASA Journal on Uncertainty Quantification}, 9(2):593--649.

\bibitem[Marelli et~al., 2021]{UQdoc_14_104}
Marelli, S., L\"uthen, N., and Sudret, B. (2021).
\newblock {UQLab user manual -- Polynomial chaos expansions}.
\newblock Technical report, Chair of Risk, Safety and Uncertainty
  Quantification, ETH Zurich, Switzerland.
\newblock Report \# UQLab-V1.4-104.

\bibitem[Marelli and Sudret, 2014]{marelli2014uqlab}
Marelli, S. and Sudret, B. (2014).
\newblock {UQLab}: A framework for uncertainty quantification in {Matlab}.
\newblock In {\em Vulnerability, Uncertainty, and Risk (Proc. 2nd Int. Conf. on
  Vulnerability, Risk Analysis and Management {(ICVRAM2014)}, Liverpool, United
  Kingdom)}, pages 2554--2563.

\bibitem[Mariethoz and Caers, 2014]{mariethoz2014multiple}
Mariethoz, G. and Caers, J. (2014).
\newblock {\em Multiple-point geostatistics: stochastic modeling with training
  images}.
\newblock John Wiley \& Sons.

\bibitem[Marzouk and Xiu, 2009]{marzouk2009stochastic}
Marzouk, Y. and Xiu, D. (2009).
\newblock A stochastic collocation approach to {B}ayesian inference in inverse
  problems.
\newblock {\em Communications In Computational Physics}, 6.

\bibitem[Marzouk et~al., 2007]{marzouk2007stochastic}
Marzouk, Y.~M., Najm, H.~N., and Rahn, L.~A. (2007).
\newblock Stochastic spectral methods for efficient {B}ayesian solution of
  inverse problems.
\newblock {\em Journal of Computational Physics}, 224(2):560--586.

\bibitem[Meles et~al., 2022]{meles2022bayesian}
Meles, G.~A., Linde, N., and Marelli, S. (2022).
\newblock Bayesian tomography with prior-knowledge-based parametrization and
  surrogate modeling.
\newblock {\em Geophysical Journal International}.

\bibitem[M{\'e}tivier et~al., 2020]{metivier2020efficient}
M{\'e}tivier, D., Vuffray, M., and Misra, S. (2020).
\newblock Efficient polynomial chaos expansion for uncertainty quantification
  in power systems.
\newblock {\em Electric Power Systems Research}, 189:106791.

\bibitem[Nagel, 2019]{nagel2019bayesian}
Nagel, J.~B. (2019).
\newblock Bayesian techniques for inverse uncertainty quantification.
\newblock {\em IBK Bericht}, 504.

\bibitem[Rasmussen, 2003]{rasmussen2003gaussian}
Rasmussen, C.~E. (2003).
\newblock Gaussian processes in machine learning.
\newblock In {\em Summer school on machine learning}, pages 63--71. Springer.

\bibitem[Reynolds et~al., 1996]{reynolds1996reparameterization}
Reynolds, A.~C., He, N., Chu, L., and Oliver, D.~S. (1996).
\newblock Reparameterization techniques for generating reservoir descriptions
  conditioned to variograms and well-test pressure data.
\newblock {\em SPE Journal}, 1(04):413--426.

\bibitem[Sacks et~al., 1989]{sacks1989designs}
Sacks, J., Schiller, S.~B., and Welch, W.~J. (1989).
\newblock Designs for computer experiments.
\newblock {\em Technometrics}, 31(1):41--47.

\bibitem[Strebelle, 2002]{strebelle2002conditional}
Strebelle, S. (2002).
\newblock Conditional simulation of complex geological structures using
  multiple-point statistics.
\newblock {\em Mathematical Geology}, 34(1):1--21.

\bibitem[Tarantola, 2005]{tarantola2005inverse}
Tarantola, A. (2005).
\newblock {\em Inverse problem theory and methods for model parameter
  estimation}.
\newblock SIAM.

\bibitem[Thibaut et~al., 2021]{thibaut2021new}
Thibaut, R., Laloy, E., and Hermans, T. (2021).
\newblock A new framework for experimental design using bayesian evidential
  learning: The case of wellhead protection area.
\newblock {\em Journal of Hydrology}, 603:126903.

\bibitem[Torre et~al., 2019]{torre2019data}
Torre, E., Marelli, S., Embrechts, P., and Sudret, B. (2019).
\newblock Data-driven polynomial chaos expansion for machine learning
  regression.
\newblock {\em Journal of Computational Physics}, 388:601--623.

\bibitem[Vrugt et~al., 2009]{vrugt2009accelerating}
Vrugt, J.~A., Ter~Braak, C., Diks, C., Robinson, B.~A., Hyman, J.~M., and
  Higdon, D. (2009).
\newblock Accelerating markov chain monte carlo simulation by differential
  evolution with self-adaptive randomized subspace sampling.
\newblock {\em International Journal of Nonlinear Sciences and Numerical
  Simulation}, 10(3):273--290.

\bibitem[Wagner et~al., 2020]{wagner2020bayesian}
Wagner, P.-R., Fahrni, R., Klippel, M., Frangi, A., and Sudret, B. (2020).
\newblock Bayesian calibration and sensitivity analysis of heat transfer models
  for fire insulation panels.
\newblock {\em Engineering Structures}, 205:110063.

\bibitem[Wagner et~al., 2021a]{wagner2021bayesian}
Wagner, P.-R., Marelli, S., and Sudret, B. (2021a).
\newblock Bayesian model inversion using stochastic spectral embedding.
\newblock {\em Journal of Computational Physics}, 436:110141.

\bibitem[Wagner et~al., 2021b]{UQdoc_14_113}
Wagner, P.-R., Nagel, J., Marelli, S., and Sudret, B. (2021b).
\newblock {UQLab user manual -- Bayesian inversion for model calibration and
  validation}.
\newblock Technical report, Chair of Risk, Safety and Uncertainty
  Quantification, ETH Zurich, Switzerland.
\newblock Report UQLab-V1.4-113.

\bibitem[Xiu and Karniadakis, 2002]{xiu2002wiener}
Xiu, D. and Karniadakis, G.~E. (2002).
\newblock The {W}iener-{A}skey polynomial chaos for stochastic differential
  equations.
\newblock {\em SIAM Journal on Scientific Computing}, 24(2):619--644.

\bibitem[Zahner et~al., 2016]{zahner2016image}
Zahner, T., Lochb{\"u}hler, T., Mariethoz, G., and Linde, N. (2016).
\newblock Image synthesis with graph cuts: a fast model proposal mechanism in
  probabilistic inversion.
\newblock {\em Geophysical Journal International}, 204(2):1179--1190.

\end{thebibliography}

\clearpage
\appendix
\begin{appendices}
\renewcommand\theequation{A.\arabic{equation}}
\setcounter{equation}{0}
\counterwithin{figure}{section}

\section{Appendix: overview of the limitations of the VAE-PCE approach}
\begin{figure}
\centering
\includegraphics[width=1\textwidth]{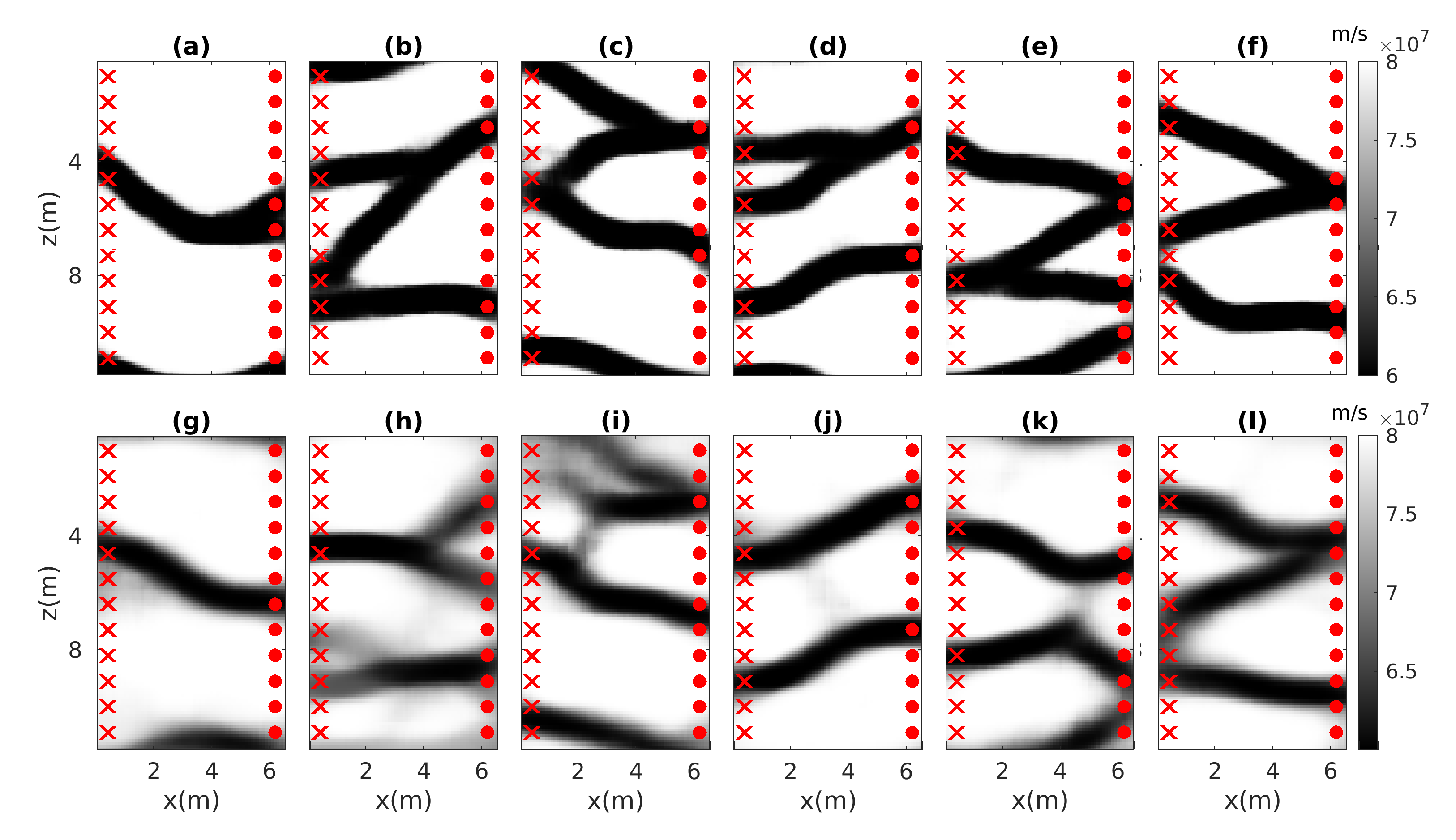}
\caption{\label{fig:APP} Reference velocity fields (a)-(f) and (g)-(l) the corresponding posterior mean images for MCMC inversion based on VAE-PCE surrogate modeling.}
\end{figure}

For the configuration considered in this manuscript, PCEs based on VAE parameters provide poor accuracy in predicting traveltimes. When calculated on a representative validation set, an aggregate rmse of $2.01$ ns is observed for the misfit between reference and predicted data.
For the velocity distribution in Fig. \ref{fig:007a}, the traveltime prediction is particularly poor, with an rmse of $3.1$ ns.
In Fig. \ref{fig:APP}, we consider six additional reference velocity fields and the corresponding posterior mean images for MCMC inversion when using VAE-PCE surrogate modeling.
In some cases the posterior mean resembles the reference velocity field well (compare \ref{fig:APP}(a) to \ref{fig:APP}(g), or \ref{fig:APP}(f) to \ref{fig:APP}(l)). However, large differences can arise between the reference and the VAE-PCE posterior mean (e.g., compare \ref{fig:APP}(b) to \ref{fig:APP}(h), or \ref{fig:APP}(e) to \ref{fig:APP}(k)). Even if the corresponding modeling error is accounted for in the inversion implying that the posterior mean models should be unbiased, we find that the modeling error has severe impacts by increasing the posterior model uncertainty.
\end{appendices}
\end{document}